\documentclass[aps,preprintnumbers,superscriptaddress,onecolumn,amsmath,amssymb,floatfix,pra]{revtex4} 
\pdfoutput=1
\usepackage{graphicx}
\usepackage[usenames,dvipsnames]{color}
\usepackage[colorlinks,bookmarks=false,citecolor=NavyBlue,linkcolor=Red,urlcolor=blue]{hyperref}
\usepackage[export]{adjustbox}

\usepackage{mathtools}

\newcommand\be{\begin{equation}}
\newcommand\bea{\begin{eqnarray}}
\newcommand\bes{\begin{subequations}}
\newcommand\esu{\end{subequations}}
\newcommand\ee{\end{equation}}
\newcommand\eea{\end{eqnarray}}

\newcommand{\cmmnt}[1]{}

\newcommand\ba         {\begin{eqnarray} } 
\newcommand\ea         {\end{eqnarray} }

\def\doi{http://dx.doi.org/}

\usepackage{braket}

\newcommand{\dd}{{\rm d}}

\usepackage{dsfont}
\begin{document}

\title{R\'enyi entanglement entropies for the compactified massless boson with open boundary conditions}
\author{Alvise Bastianello}
\affiliation{Institute for Theoretical Physics, University of Amsterdam, Science Park 904, 1098 XH Amsterdam, The Netherlands}

\graphicspath{{figures/}}

\date{\today}


\begin{abstract}
We investigate the R\'enyi entanglement entropies for the one-dimensional massless free boson compactified on a circle, which describes the low energy sector of several interacting many-body 1d systems (Luttinger Liquid). We focus on systems on a finite segment with open boundary conditions and possible inhomogeneities in the couplings. We provide expressions for the R\'enyi entropies of integer indices in terms of Fredholm determinant-like expressions. Within the homogeneous case, we reduce the problem to the solution of linear integral equations and the computation of Riemann Theta functions. We mainly focus on a single interval in the middle of the system, but results for generic bipartitions are given as well.
\end{abstract}

\maketitle

\section{Introduction}

The challenge of quantifying the entanglement content of extended systems is at the common intersection of several communities, ranging from high energy to quantum information and condensed matter \cite{rev-enta,rev-enta2,rev-enta3}, with a recent emphasis on out-of-equilibrium features (see Ref. \cite{CaCa16} and references therein). 
It has been understood that different phases of many-body systems can be characterized in terms of their entanglement properties \cite{CaCa04,CaCa09}, which also cover an important role in the efficiency of tensor networks algorithms \cite{swvc-08,swvc-082,pv-08,rev0,d-17,lpb-18}.
Furthermore, recent times have witnessed a direct engineering of entanglement measurements in cold atom experiments \cite{Ra15,Kauf16,DaPi12,BrEl19,LuRi19,ElVe18,UnZh17}.

Given a quantum system initialized in a pure state $|\psi\rangle$ and bipartited in two halves $A\cup B$, arguably the most widespread measurement of the entanglement between $A$ and $B$ is given by the Von Neumann entropy
\be
\mathcal{S}_{B}=-\text{Tr}_B \hat{\rho}_{B}\log\hat{\rho}_B\, ,
\ee
where $\hat{\rho}_B$ is the reduced density matrix of the subsystem $B$, obtained through a partial trace of the degrees of freedom of the subsystem $A$, i.e. $\hat{\rho}_B\equiv\text{Tr}_A |\psi\rangle\langle \psi|$. Together with the Von Neumann entanglement entropies, one can consider the more general $N^\text{th}$ R\'enyi entropy, defined as
\be
\mathcal{S}_{B}^N=\frac{1}{1-N}\log \text{Tr}_B \hat{\rho}_B^N\, .
\ee
The knowledge of all the moments of the reduced density matrix gives information about its whole spectrum \cite{CaLe08,AlCaTo18,LiHa08}: the R\'enyi entropies are the central objects of this work.
The Von Neumann entanglement entropy can be understood as the limit $N\to 1$ of the R\'enyi entropies. Such a point of view is often used as a useful computational tool, commonly known as replica trick \cite{CaCa04}: the R\'enyi entropies for integer indices $N$ are computed, subsequently the result is analytically continued to real indices and the limit $N\to 1$ is taken.
For generic systems, the computation of R\'enyi entropies is a tremendous task on its own, not to speak of their analytical continuation. A great insight has been achieved at criticality: the system acquires scale invariance and the powerful methods of conformal field theory (CFT) \cite{BPZ,difrancesco} can be used to investigate the scaling part of the entanglement entropies. In the one dimensional world, CFT displays its full power and allows one to derive an array of simple analytical results. For example, for an infinite system in the ground state with a bipartition chosen to be an interval of length $L$ and the rest, the following result holds true \cite{CaCa04,CaCa09,HoLaWi94}
\be\label{eq_S_infinity}
\mathcal{S}^N_L \, = \, c\frac{1+1/N}{6} \log \left( \frac{L}{a_N}  \right)  \, .
\ee
Above, $c$ is the central charge of the CFT and corrections to the conformal prediction go to zero as $L\to\infty$. The constant $a_N$ (but in principle $N-$dependent) is an ultraviolet cutoff determined by the microscopic features of the model (i.e. not fixed by the CFT). If the microscopic system lives on a lattice, then $a_N$ si proportional to the lattice spacing.
Several many-body quantum systems can show critical features, in particular the compactified free boson (Luttinger Liquid) \cite{Ha81,Ha81bis,Ha81tris,Ca04,Gi04,Ts07,CaCiGiOrRi11} is a ubiquitous low-energy description of models with a $U(1)$ conserved charge, such as the total number of particles in cold atoms systems or the total magnetization in spin chains.
The compactified boson is arguably one of the simplest CFT, having the central charge equal to one $c=1$ \cite{Ca04,Gi04}.

While it remains true that the scaling part of the entanglement entropies is determined by the CFT, on the other hand the global conformal invariance alone does not uniquely fix the entanglement of arbitrary bipartitions, which in turn depends on the details of the underlying CFT. For example, in a system with periodic boundary conditions (PBC) and a bipartition made out of two disjoint intervals with the rest, the global conformal symmetry is not enough to determine the entanglement entropy. A lot of efforts have been directed towards this problem \cite{CaCa04,RuToCa18,RaGl12,CaGl08,FuPaSh09,He10,HeLaRo13,
CaHu09bis,FaFlInPa08,AlTaCa10,AlTaCa11,IgPe10,FaCa10,Ca10,Fa12,
ChLoZh13,ChLoZh14,ArEsFa14,CoToCa16,LiZh16,LiuLiu16,BeKeZa17,
MuMu18,DuEsIk18,XaAlSi18}, obtaining closed analytical expressions for the compactified boson \cite{CaCaTo09,CaCaTo11,CoTaTo14}.

Another situation where one cannot rely just on the global conformal symmetry is finite subsystems on an interval $[0,L]$, with some boundary conditions.
Apart from the per se interest, the study of these boundary CFTs \cite{Ca08} has many applications going from Kondo physics \cite{Ko64,He97,AfLaSo09} to string theory. Indeed, the entanglement in Kondo problems has been already intensively studied (see e.g. Ref. \cite{rev-enta3} and references therein). In this paper, we start analyzing this problem from the simplest possible case, which are open boundary conditions (OBC).

Let us consider a bipartition $A\cup B$ where we choose $B=[x_1,x_2]$ with $x_1<x_2$: if one of the two extrema coincides with the boundaries of the system, i.e. $x_1=0$ or $x_2=L$, conformal invariance unambiguously predicts the entanglement entropy \cite{CaCa04,CaCa09} (albeit the $\mathcal{O}(L^0$) contribution can acquire an extra term due to the non trivial boundaries \cite{CaCa04,ZhBaFj06}, which is the boundary entropy of Affleck and Ludwig \cite{AfLu91}).
On the other hand, if $0<x_1< x_2<L$, global conformal symmetry is not sufficient anymore and a detailed analysis of the operator content of the CFT is required. 
In this case, closed analytical results are known only for free fermionic systems \cite{CaFoHu05,FaCa11,CaMiVi11,CaMiVi11JS}.

OBC are of primary importance to be explored, both for the experimental relevance and for their importance in numerical studies.
Indeed, the advent of DMRG techniques \cite{ScDMRGrev} provided us a powerful numerical tool to study the one-dimensional world and these are best suited to work with OBC.
However, we stress that while the entanglement entropies of bipartitions starting at the boundaries  (i.e. choosing $A\cup B$ with $A=[0,\bar{x}]$) are easily computable with DMRG methods, studying the case of an interval in the middle of the system in the scaling regime is much harder. In this perspective, alternative analytical and semi-analytical tools are surely helpful.

Going beyond the global conformal invariance of the theory is extremely hard, but dealing with the compactified boson we can take advantage of an important simplification: the model is free.
Efficient techniques exist to deal with free (i.e. gaussian) systems \cite{CaHu09}, and even though extracting the scaling part of the entanglement in terms of simple analytical expressions can be hard, the result is straightforwardly numerically computed diagonalizing matrices which scale proportionally with the system size. Such a task can be easily performed on a laptop for quite large sizes, much faster compared with extracting the same result with DMRG methods.
However, the techniques of Ref. \cite{CaHu09} cannot be straightforwardly applied to the compact boson because of the non-trivial compactification radius, which does affect the entanglement entropy (see for instance Ref. \cite{CaCaTo09,CaCaTo11,CoTaTo14}).
In this work, we address the problem of computing the R\'enyi entropies for integer indices on the compact free boson: our final goal are simple semi-analytic expressions which can be regarded as a generalization of the free-model computations of Ref. \cite{CaHu09}, but properly taking in account the compactification radius.

Recent times witnessed an increasing interest towards inhomogeneous generalizations of the compact boson, due to their capability of describing the low energy sector of inhomogeneous systems \cite{DuStViCa17,DuStCa17,BrDu18,BrDu17,EiBa17,RuBrDu19,MuRuCa19}. In this perspective, we consider the following more general inhomogeneous Hamiltonian \cite{BrDu18} 
\be\label{eq_inh_LL}
\hat{H}^{\text{LL}}=\frac{1}{2\pi}\int_0^L \dd x \, v(x)\,\left[  K(x)  (\partial_x \hat{\theta})^2+\frac{1}{K(x)} (\partial_x\hat{\phi} )^2\right]\, ,
\ee
where the phase and density fields (respectively $\hat{\theta}$ and $\hat{\phi}$) are canonically conjugated $[\hat{\theta} (x),\partial_y\hat{\phi}(y)]=\pi i\delta(x-y)$. The model is free, but the target space of the phase field is compactified
\be\label{eq_cR}
\hat{\theta}(x)\equiv \hat{\theta}(x)+2\pi\, .
\ee
The density field is also compactified, but modulus $\pi$.
Choosing in Eq. \eqref{eq_inh_LL} the sound velocity $v(x)>0$ and the Luttinger parameter $K(x)>0$ to be constant, the usual homogeneous case is recovered. It must be stressed that while the presence of an inhomogeneous velocity does not spoil conformal invariance \cite{DuStViCa17,DuStCa17}, an inhomogeneous $K(x)$ does: therefore it is of utmost importance to have other efficient methods to access entanglement entropies which do not rely on CFT.
The parameters $v$ and $K$ are model-dependent and must be properly fixed by an independent analysis of the microscopic system at hand: for example, $v$ and $K$ can be extracted in integrable models by means of Bethe Ansatz techniques \cite{Gi04,Ca04,BrDu18}, or numerically in generic systems.

As anticipated, for the sake of simplicity we focus on finite systems with OBC, but we expect that the same techniques can be applied to different boundary conditions as well.
OBC on the microscopic system impose Dirichlet boundary conditions on the density field of the Luttinger liquid \cite{Ca04}
\be\label{eq_field_boundary}
\hat{\phi}(x=0)=\phi_\text{left}\,,\hspace{2pc}\hat{\phi}(x=L)=\phi_\text{right}\, ,
\ee
while the phase fluctuation at the boundaries is left free. 
The conservation of the underlying $U(1)$ symmetry of the microscopic model imposes $\phi_\text{left}-\phi_\text{right}\in \pi \mathbb{Z}$  \cite{Ca04}.

We will see that, once the difference in the boundary conditions is fixed to a multiple of $\pi$, the R\'enyi entropies are otherwise independent on the actual choice of the boundary fields. 

We are primarily interested in a bipartition $A\cup B$ with $B=[x_1,x_2]$ and $0<x_1< x_2<L$, but we also present results for arbitrary bipartitions.
Within the fully inhomogeneous case, we provide general expressions in terms of Fredholm determinants.
Then, we further advance within the homogeneous case, reducing the problem to a solution of linear integral equations and to the computation of Riemann Theta functions, analogously to the findings of Ref. \cite{CaCaTo09,CaCaTo11,CoTaTo14}. These operations can be quickly carried out on a laptop.

For the sake of clarity, we anticipate and discuss our results in the forthcoming Section \ref{sec_result}: we  present the general expression valid also within the inhomogeneous case and then consider its simplified version for the homogeneous system.
The technical derivation is contained in Section \ref{sec_derivation} and then our conclusions are gathered in Section \ref{sec_conclusion}.
A short description of the numerical methods is given in Appendix \ref{app_latt}.

\section{Summary of the results}
\label{sec_result}

In this Section we present and discuss our main result, leaving the most cumbersome derivations to Section \ref{sec_derivation}. 
As already stated in the introduction, we are mostly interested in the case of a bipartition $A\cup B$ with $B=[x_1,x_2]$ and $0<x_1<x_2<L$. 
Generalizations to arbitrary bipartitions are possible as well and are presented in Section \ref{sec_gen_bip}.
In the case of a single interval in the middle of the system, the $N^\text{th}$ R\'enyi entanglement entropy is given by the following expression
\be\label{eq_result}
\mathcal{S}_{[x_1,x_2]}^N=\frac{1}{1-N}\log\left[\prod_{\ell=1}^{N-1}\sqrt{\frac{1}{\pi^3\mathcal{I}_{\ell/N}}\frac{1}{\det\left(\frac{1+\Phi^{-1}\Phi_{\ell/N}}{2}\right)}}\sum_{\{m_j\}_{j=1}^{N-1}}  \exp\Bigg\{-4\sum_{a,b=1}^{N-1}\mathcal{M}_{ab}\, m_a m_b\Bigg\}\, \right]+\log g\, .
\ee

Above, we introduced the following notation:
\begin{enumerate}
\item $\Phi$ is a linear operator constructed from the two-point connected correlator of the density field $\Phi(x,y)=\langle \hat{\phi}(x)\hat{\phi}(y)\rangle-\langle \hat{\phi}(x)\rangle\langle\hat{\phi}(y)\rangle$. We then define $\Phi_{\Omega}$ starting from $\Phi$ and adding a twist with a phase $2\pi \Omega$ at the boundaries of $B$ 
\be\label{eq_phiomega}
\Phi_{\Omega}(x,y)= \Phi(x,y) e^{i2\pi \Omega (\chi_B(y)-\chi_B(x))}\, ,
\ee
with $\chi_{B}(x)$ the characteristic function of the subsystem $B$, being $1$ if $x\in B$ and zero otherwise.
\item $\det\left(\frac{1+\Phi^{-1}\Phi_{\ell/N}}{2}\right)$ is understood as a Fredholm determinant, being the domain of the operator the whole system $[0,L]$. From a practical point of view, the two-point correlator is discretized on a proper lattice and the determinants are computed, then the limit of infinitesimal lattice spacing is considered (see Appendix \ref{app_latt}). While taking this limit, the logarithm of the product of determinants is well defined up to a constant, which logarithmically diverges in the limit of zero lattice size and it is due to the twist in Eq. \eqref{eq_phiomega}. Such a divergence is ubiquitous in entanglement's studies (see e.g. \cite{CaCa04,CaCa09}). Indeed, the lattice discretization introduces a UV cutoff which affects the entanglement according to Eq. \eqref{eq_S_infinity} (see Appendix \ref{app_latt}).
\item The summation in Eq. \eqref{eq_result} is over all the possible integers $m_a\in \mathbb{Z}$ and is an example of a Riemann Theta function. Its appearance is due to the non-trivial compactification radius and similar functions are known to be present in the R\'enyi entropy with PBC \cite{CaCaTo09,CaCaTo11,CoTaTo14} as well.
\item $g$ is the ground state degeneracy, giving rise to the Affleck-Boundary boundary entanglement \cite{AfLu91,ZhBaFj06}. Within our approach, $g$ emerges as an ill-defined constant (i.e. independent on $x_1$ and $x_2$) whose computation requires a proper regularization scheme yet to be devised.
For the homogeneous compact boson with open boundary conditions, i.e. Dirichlet boundary conditions on the density field, the ground state degeneracy has been computed in Ref. \cite{OshAf97} with other methods, as we further comment in the next section.

\item The matrix $\mathcal{M}$ is defined as
\be\label{eq_M_def}
\mathcal{M}_{ab}=\sum_{\ell=1}^{N-1} \frac{e^{-i2\pi \ell(a-b)/N}}{N\mathcal{I}_{\ell/N}}\, 
\ee
and lastly the real coefficients $\mathcal{I}_{\ell/N}$ are obtained from the solution of a linear integral equation
\be\label{eq_Il_def}
\mathcal{I}_{\ell/N}=\int_{x_1}^{x_2}\dd x\, s_{\ell/N}(x)\, , \hspace{4pc}\int_0^L\dd y\,(\Phi(x,y)+\Phi_{\ell/N}(x,y))s_{\ell/N}(y)=\chi_B(x)\, .
\ee
The solution $s_{\ell/N}(x)$ displays a power-law singularity near the endpoint $x_1$
\be\label{eq_pow_s}
s_{\ell/N}(x)\sim \left|\frac{1}{x-x_1}\right|^{\mu(\ell/N)}\, ,
\ee
and similarly at $x_2$. 
We were not able to analytically characterize such a divergence, but the numerical tests provided in Appendix \ref{app_latt} confirm the power-law singularity and set $0.5\le \mu(\ell/N)\le 1$, with $\mu$ being $1$ only at $\ell=N$ (indeed, in the homogeneous case and $\ell=N$, Eq. \eqref{eq_Il_def} can be explicitly solved and the power-law divergence is analytically confirmed).
Note that away from the single point $\ell=N$, the power-law singularity \eqref{eq_pow_s} is integrable, thus $\mathcal{I}_{\ell/N}$ is finite.
Therefore, the R\'enyi entropy is well defined except for the overall UV divergent offset caused by the Fredholm determinants, which we have already discussed.
\item We stress that the connected correlator $\Phi(x,y)$ is independent from the precise choice of the boundary fields in Eq. \eqref{eq_field_boundary}, which therefore do not affect the R\'enyi entropies.
\end{enumerate}

Eq. \eqref{eq_result} requires as an input the two-point connected correlation function of the density field. For the general case where both the sound velocity and the Luttinger parameter are space dependent, the correlator can be determined solving a two dimensional electrostatic problem \cite{BrDu18}. 
The two-point function is efficiently numerically computed, but a general analytical solution is unknown. If $K(x)=\text{const.}$, a putative space-dependent velocity can be absorbed in a change of coordinates
\be
x\to x'=\int_0^x\dd y\, \frac{1}{v(y)}\, ,
\ee
along the lines of Ref. \cite{DuStViCa17}. In the new coordinates, the sound velocity is simply $v=1$ and the homogeneous problem is easily solved \cite{Ca04}.
As we already commented, Eq. \eqref{eq_result} can be generalized to bipartitions in an arbitrary number of intervals: the structure of the expression remains the same, but $\mathcal{I}_{\ell/N}$ becomes a matrix of dimension $(N_{I}-2)\times (N_{I}-2)$ with $N_{I}$ the number of intervals of the bipartition (note that for $N_{I}=3$ we retrieve the previous result of a single interval in the middle: indeed the dimension of the matrix becomes one). The matrix $\mathcal{M}_{ab}$ is promoted to a tensor $\mathcal{M}_{ab}\to \mathcal{M}_{ab}^{ij}$ where lower indices live in a $(N-1)-$dimensional space and upper indices in a space of dimension $(N_{I}-2)$. Therefore, the Riemann Theta function requires a summation over $(N-1)\times(N_{I}-2)$ independent integers.
We provide the explicit expression, together with its derivation, in Section \ref{sec_gen_bip}.
We now specialize our findings to the homogeneous case, providing a benchmark with known results and further simplifying the expression.

\subsection{The homogeneous case}
\label{sec_hom_ren}

We now revert to the homogeneous case, i.e. we consider $K$ and $v$ to be constant along the system.
The homogeneous case stands out as one of the most interesting applications of our approach, where it can be compared with CFT results.
As we have already commented, CFT gives important insight into the analytic form of the entanglement entropies, albeit it is not able to completely determine it. Exact, close results are known only at the free fermion point \cite{CaFoHu05,FaCa11,CaMiVi11,CaMiVi11JS}.
Here, we greatly exploit the available information: first of all, the free fermion result provides a non trivial check of the correctness of our computation. Furthermore, by a direct comparison of the general form of the CFT expression with our result Eq. \eqref{eq_result}, we can greatly simplify the latter.

For the case of a single interval, global conformal invariance fixes the entanglement entropy up to a function of the four-point ratio $X$, constructed from the boundaries and the extrema of the interval \cite{CaCa09,CaCaTo09,CaCaTo11}
\be\label{eq_S_free_fermion}
\mathcal{S}_{[x_1,x_2]}^N=\frac{N+1}{12N}\log\left[\frac{2L}{\pi}\frac{\sin\left(\frac{\pi}{2L} |x_1-x_2|\right)\sin\left(\frac{\pi}{L}x_1\right)\sin\left(\frac{\pi}{L}x_2\right)}{\sin\left(\frac{\pi}{2L} (x_1+x_2)\right)}\right]+\frac{1}{1-N}\log\mathcal{F}_N(X)+\log g+\text{const.}\, ,\,\,\,\, X=\frac{x_1(L-x_2)}{L(x_2-x_1)}\, .
\ee
Above, the constant part is due to the microscopic UV cutoff.
The result can be generalized to arbitrary bipartitions. 
The function $\mathcal{F}_N$ is universal, in the sense that is fully determined by the underlying CFT and not by the microscopic features of the model, which contribute as a constant UV-divergent offset.

The function $\mathcal{F}_N$ for OBC is so far known only at the free fermion point $K=1$, where $\mathcal{F}_N=1$ \cite{CaFoHu05,FaCa11,CaMiVi11,CaMiVi11JS}  (see also Ref. \cite{DuStViCa17}): we can use the free fermion result as a non trivial check of our general expression Eq. \eqref{eq_result}. Furthermore, such a comparison allows us to derive a compact expression for $\mathcal{F}_N$ for arbitrary values of the Luttinger parameter $K$.
The ground state degeneracy for the compactified boson with Dirichlet boundary conditions has been computed in Ref. \cite{OshAf97} (see also Ref. \cite{Af98}), leading to the simple expression
\be\label{eq_hom_g}
g=K^{1/4}\, .
\ee
\begin{figure}[t!]
\begin{center}
\includegraphics[width=0.5\textwidth]{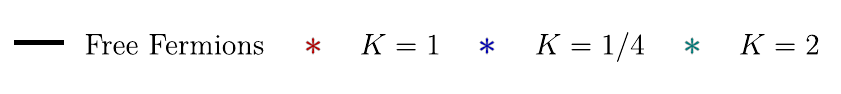}\\
\end{center}
\includegraphics[width=0.3\textwidth]{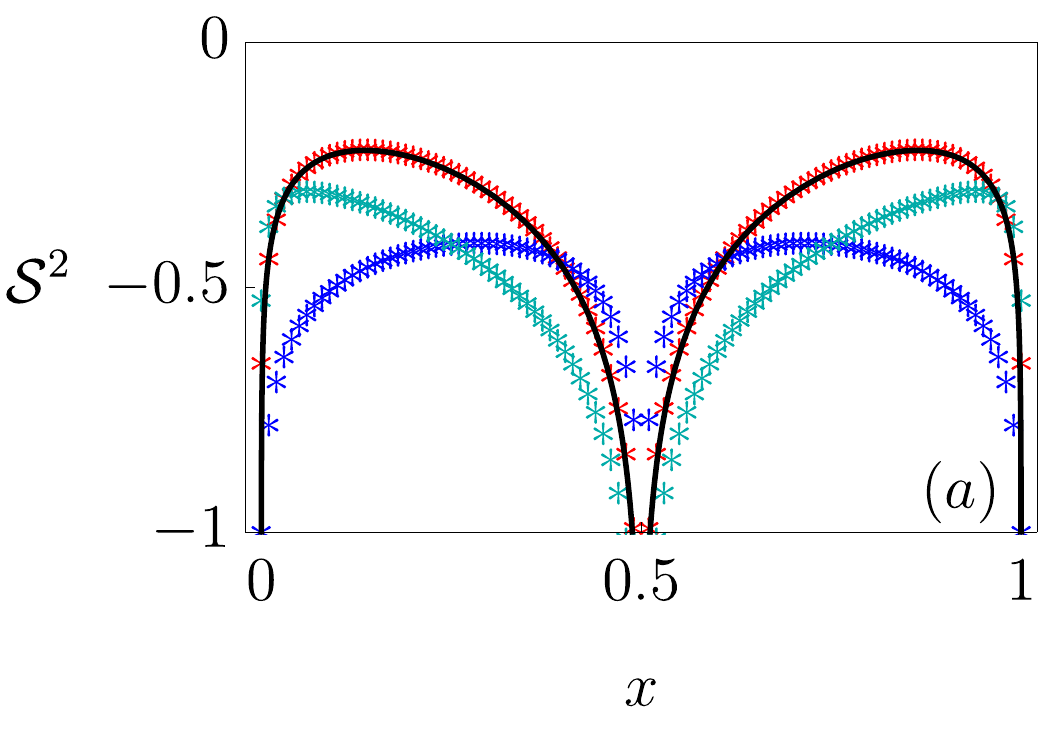}
\includegraphics[width=0.3\textwidth]{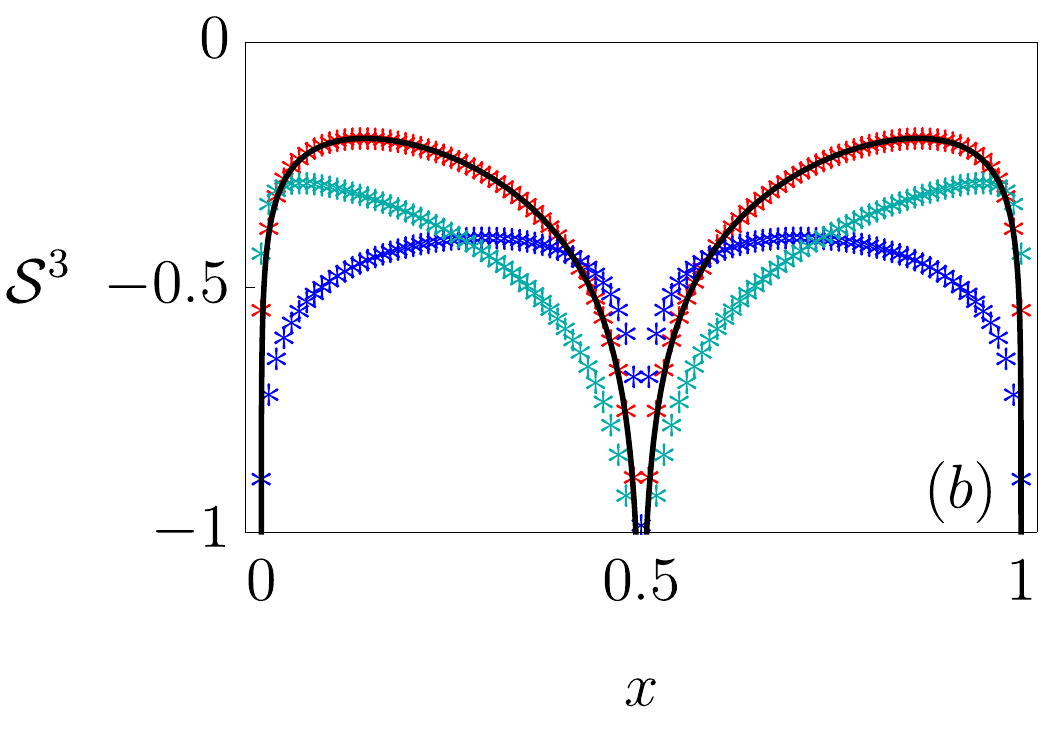}
\includegraphics[width=0.3\textwidth]{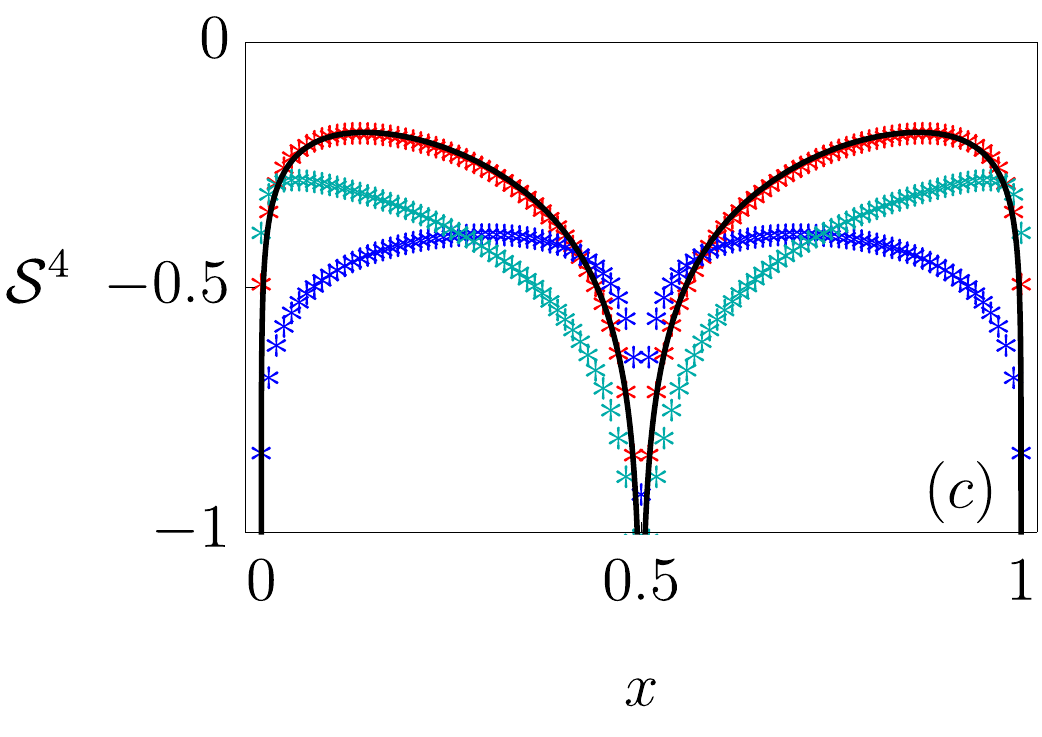}\\
\includegraphics[width=0.3\textwidth]{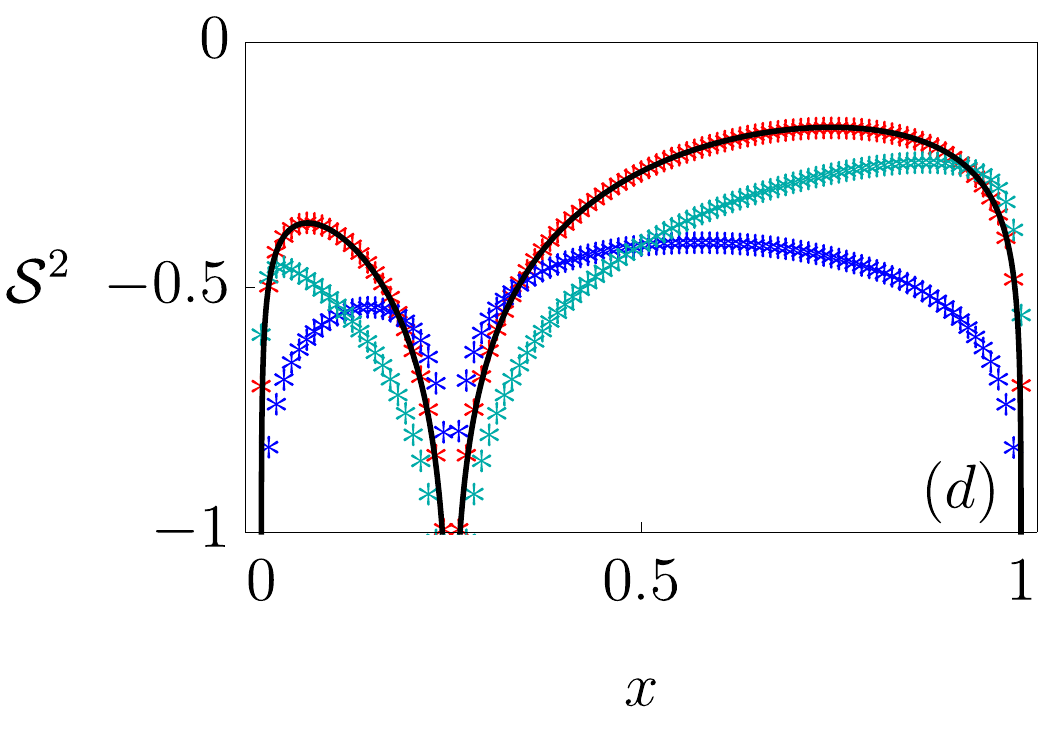}
\includegraphics[width=0.3\textwidth]{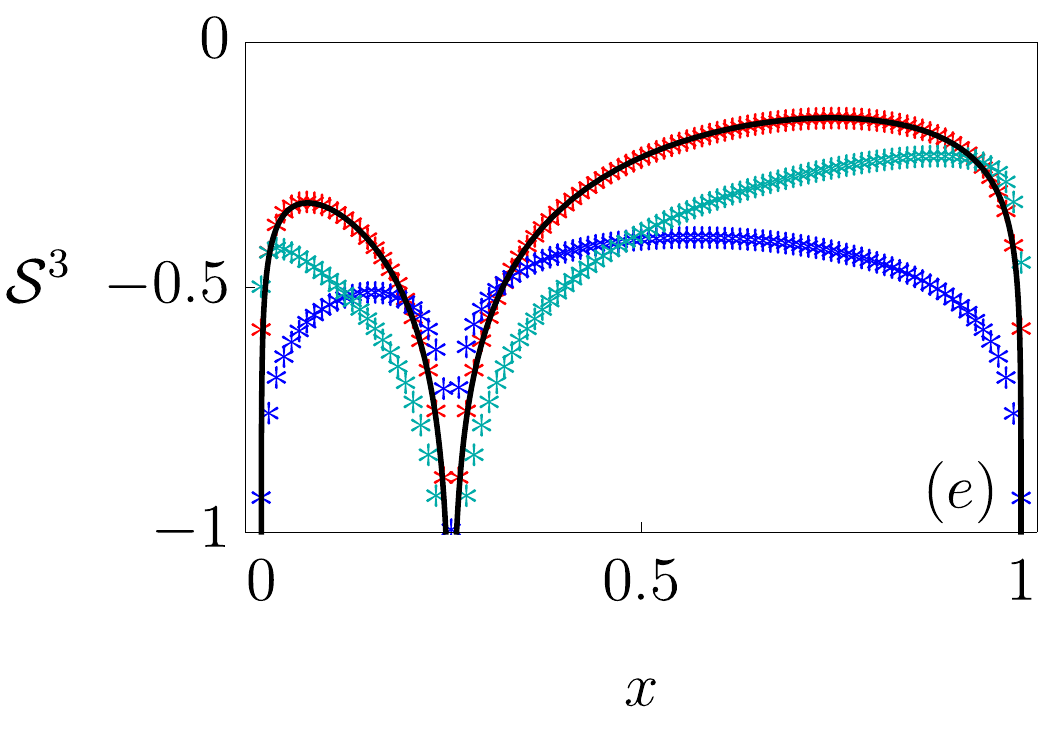}
\includegraphics[width=0.3\textwidth]{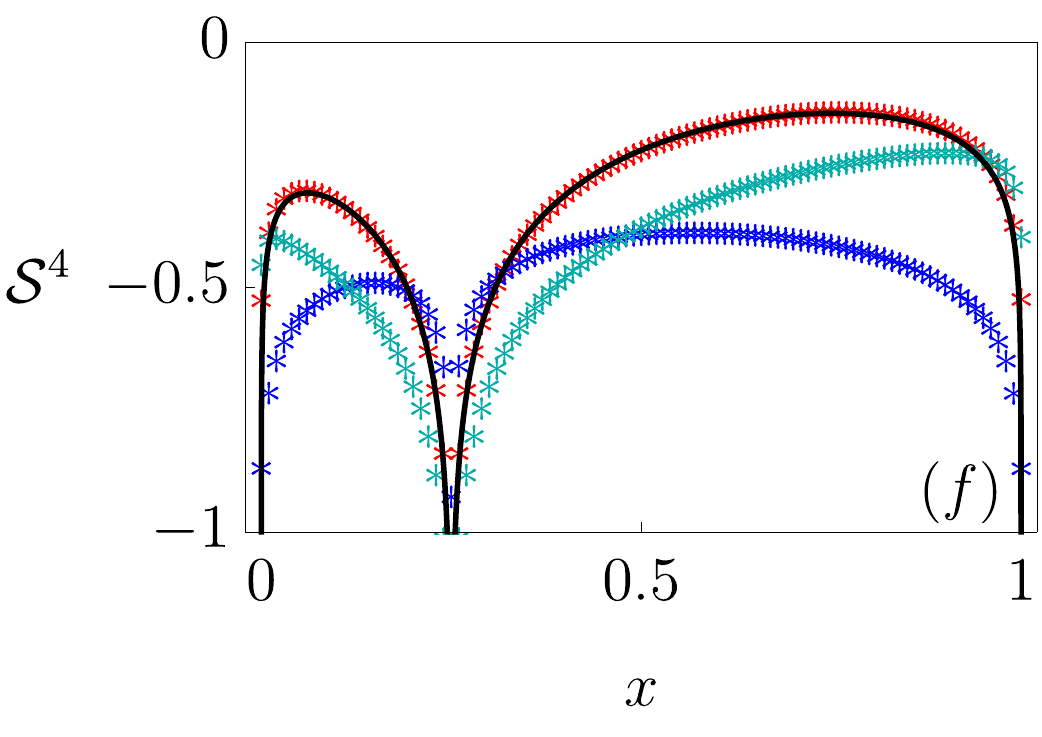}
\caption{\label{fig_1}\emph{We numerically evaluate Eq. \eqref{eq_result}, according to the lattice discretization discussed in Appendix \ref{app_latt}, for different values of the Luttinger parameter $K$ and compare it against the free fermion result \eqref{eq_S_free_fermion}, which must be reproduced for $K=1$. For the ground state degeneracy $g$, we employ the expression Eq. \eqref{eq_hom_g}. Subfigs. $(a)$, $(b)$, and $(c)$: we consider the R\'enyi entropies of indices $N=\{2,3,4\}$ respectively, the system size is chosen to have unitary length $L=1$ and in the bipartition $A\cup B$ we choose $B=[0.5,x]$, plotting the entropies as a function of $x$. Subfigs. $(d)$, $(e)$, and $(f)$: same plots, but with $B=[0.25,x]$.
In evaluating Eq. \eqref{eq_result} on a discrete lattice, we confirmed the expected UV divergent constant offset (independent from $K$): in each plot we shift the numerical curves with a constant offset which guarantees the agreement with the free fermion analytical result at $K=1$. Note, in contrast with the PBC results \cite{CaCaTo09,CaCaTo11}, the break down of the $K\to 1/(4K)$ duality due to the open boundary conditions.
}}
\end{figure}
Let us now revert to Eq. \eqref{eq_result}, more specifically we consider the two-point correlator of the density field. In the homogeneous case we have the following simple expression \cite{Ca04}

\be\label{eq_corr_hom}
\Phi(x,y)= K\times[f(x-y)-f(x+y)]\, , \hspace{2pc} f(x)=-\frac{1}{4}\log\left[\sin^2\left(\frac{\pi}{2L}x \right)\right]\, .
\ee
The Luttinger parameter can be simply factorized out in the two-point correlator (this can actually be seen directly from the Hamiltonian \eqref{eq_inh_LL}, where a homogeneous Luttinger parameter can be absorbed in a rescaling of the fields).
Establishing analytically the equivalence between Eq. \eqref{eq_result} and the CFT result is extremely hard (see e.g. \cite{FaCa11,CaMiVi11,CaMiVi11JS} for a closely related problem), but numerical tests can be easily performed.
In Fig. \ref{fig_1} we plot Eq. \eqref{eq_result} for different bipartitions and different Luttinger parameters, finding perfect agreement with the known result at the free fermion point.
We can now take advantage of the free fermion limit to further simplify the general result (within the homogeneous case).
It is convenient to make $K$ explicit in Eq. \eqref{eq_result} and define the rescaled correlator
\be
\bar{\Phi}(x,y)=f(x-y)-f(x+y)\, .
\ee
Using $\bar{\Phi}$ rather than $\Phi$ in Eq. \eqref{eq_result}, we note that $K$ cancels out from the ratio of Fredholm determinants, having a non-trivial contribution only in the Riemann Theta function. Comparing with Eq. \eqref{eq_S_free_fermion} and imposing that for $K=1$ the free fermion result is recovered, we conclude
\begin{multline}\label{eq_16}
\mathcal{S}_{[x_1,x_2]}^N=\frac{N+1}{12N}\log\left[\frac{2L}{\pi}\frac{\sin\left(\frac{\pi}{2L} |x_1-x_2|\right)\sin\left(\frac{\pi}{L}x_1\right)\sin\left(\frac{\pi}{L}x_2\right)}{\sin\left(\frac{\pi}{2L} (x_1+x_2)\right)}\right]+\log g\\
\frac{1}{1-N}\log\left[K^{(N-1)/2}\frac{\sum_{\{m_j\}_{j=1}^{N-1}}  \exp\Bigg\{-4K\sum_{a,b=1}^{N-1}\bar{\mathcal{M}}_{ab}\, m_a m_b\Bigg\}}{\sum_{\{m_j\}_{j=1}^{N-1}}  \exp\Bigg\{-4\sum_{a,b=1}^{N-1}\bar{\mathcal{M}}_{ab}\, m_a m_b\Bigg\}}\, \right]+\text{const.}
\end{multline}
where $\bar{\mathcal{M}}$ is defined as per $\mathcal{M}$, but replacing $\Phi\to\bar{\Phi}$ in all the expressions.

Once again, we stress that the constant term does not depend either on the position of the interval, or on the value of $K$.
Comparing the above with Eq. \eqref{eq_S_free_fermion}, we can express the universal function $\mathcal{F}_N$ as a ratio of Riemann Theta functions
\be\label{eq_FN}
\mathcal{F}_N(X)=K^{(N-1)/2}\frac{\sum_{\{m_j\}_{j=1}^{N-1}}  \exp\Bigg\{-4K\sum_{a,b=1}^{N-1}\bar{\mathcal{M}}_{ab}\, m_a m_b\Bigg\}}{\sum_{\{m_j\}_{j=1}^{N-1}}  \exp\Bigg\{-4\sum_{a,b=1}^{N-1}\bar{\mathcal{M}}_{ab}\, m_a m_b\Bigg\}}\, .
\ee
\begin{figure}[t!]
\includegraphics[width=0.45\textwidth]{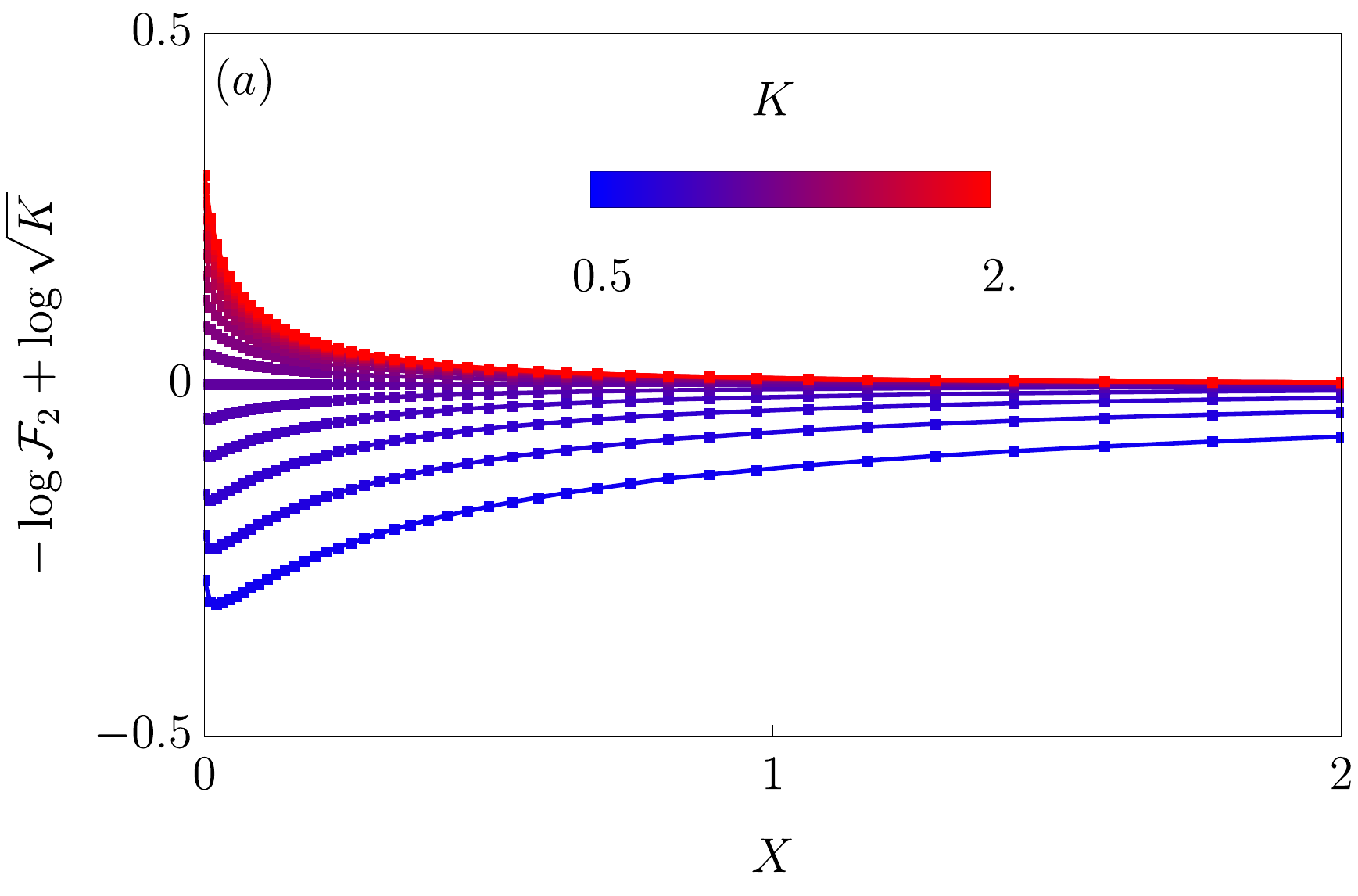}
\hspace{1pc}
\includegraphics[width=0.45\textwidth]{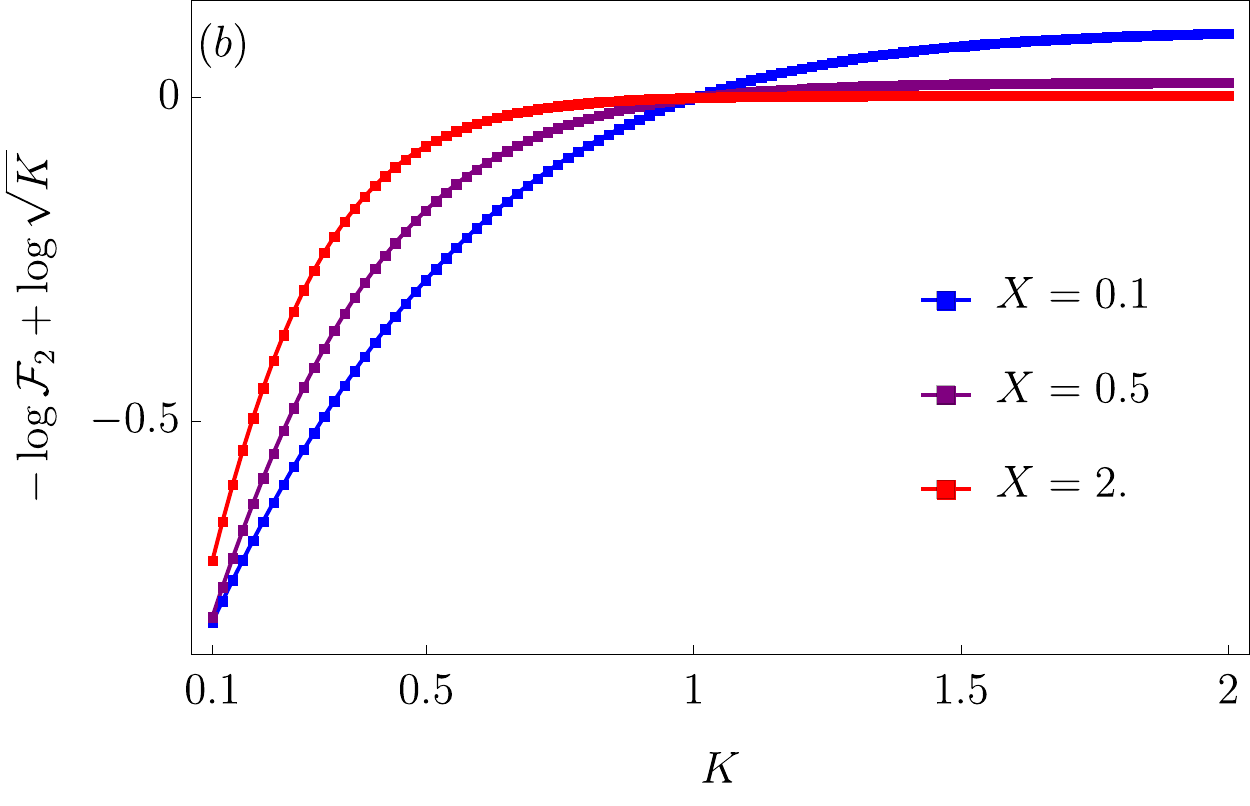}
\caption{\label{fig_FN}\emph{
We plot the universal function $\mathcal{F}_N(X)$ \eqref{eq_FN} for the first non trivial R\'enyi index $N=2$. We consider its logarithm and subtract a term proportional to $\log \sqrt{K}$ in order to emphasize the role played by the Riemann Theta functions.
Subfig. $(a)$: $\mathcal{F}_N(X)$ is plotted as a function of $X$ for several choices of $K$. For $K=1$,  we obviously get $\log\mathcal{F}_2(X)=0$.
Subfig. $(b)$:  $\mathcal{F}_N(X)$ is plotted at fixed values of $X$ as a function of $K$.
}}
\end{figure}

We have numerically checked that $\mathcal{F}_N$ is a function of the four-point ratio comparing the r.h.s. of the above expression for several choices of the interval, but having the same four-point ratio. 
In Fig. \ref{fig_FN} we focus on the universal function $\mathcal{F}_N$ for R\'enyi index $N=2$. 

While the lack of a fully analytical expression for $\mathcal{M}$ makes difficult to grasp the behavior in $X$ directly from Eq. \eqref{eq_FN}, the trend in $K$ at fixed $X$ is easily understood.
In $\mathcal{F}_{N>1}$, for large $K$ the Riemann Theta function at the numerator exponentially approaches $1$, leaving out the algebraic dependence given by the prefactor $K^{(N-1)/2}$.
For small $K$ we rather reach a plateau, indeed we can readily estimate 
\be
\sum_{\{m_j\}_{j=1}^{N-1}}  \exp\Bigg\{-4K\sum_{a,b=1}^{N-1}\bar{\mathcal{M}}_{ab}\, m_a m_b\Bigg\}\simeq \int \dd^{N-1} x\, \exp\Bigg\{-4K\sum_{a,b=1}^{N-1}\bar{\mathcal{M}}_{ab}\, x_a x_b\Bigg\}=\left(\frac{\pi}{4K}\right)^{(N-1)/2}\frac{1}{\sqrt{\det \bar{\mathcal{M}}}}.
\ee
The $K-$divergence above is exactly compensated by the prefactor in Eq. \eqref{eq_FN}, leading to a finite result in the decompactification limit.

It is worth mentioning an important difference with Ref. \cite{CaCaTo09,CaCaTo11}, where the universal function $\mathcal{F}_N$ for two intervals in a system with PBC is presented. Similarly to the OBC case at hand, also the PBC result can be written in terms of Riemann Theta functions, but the expression is symmetric for $K\to 1/(4K)$.
Indeed, the compact boson with PBC presents such a duality \cite{Ca04}, which is then reflected into the entanglement entropy. 

The weak-strong coupling duality can be understood to be equivalent to swapping the phase and density fields: periodic boundary conditions are of course invariant under such an operation and the model is dual.
On the contrary, in the OBC case this is no longer true, since the density and phase fields have Dirichlet and free boundary conditions respectively, thus the duality is spoiled. Indeed, Eq. \eqref{eq_FN} is not symmetric for $K\to 1/(4K)$. 
The breakdown of the duality is clearly displayed both in Fig. \ref{fig_1} and Fig. \ref{fig_FN}.

\section{Derivation of the results}
\label{sec_derivation}

This section gathers all the technical steps which allowed us to compute Eq. \eqref{eq_result}. Before going deeper into the technical analysis, it is convenient to sketch the main steps. We compute the R\'enyi entropies through a brute-force evaluation of $\text{Tr}\hat{\rho}_B^N$ in a path integral formalism.
In this perspective, we introduce the coherent states for the phase-density field as the eigenvectors of the operator $\hat{\theta}(x)$
\be
\hat{\theta}(x)\ket{\theta}=\theta(x)\ket{\theta}\, .
\ee
Above, $\theta(x)$ is just a number.
The coherent states are complete and we explicitly perform partial traces and products in this basis.
Let $|GS\rangle$ be the ground state of the compact boson, we define the wavefunction $W[\theta]$ as its projection on the coherent state $|\theta\rangle$
\be
W[\theta]=\langle \theta|GS\rangle\, .
\ee

A common representation of this wavefunction is that of a path integral in Euclidean time \cite{CaCa04,CaCa09}, leading to a representation of the R\'enyi entropies in terms of a partition function on an $N-$sheeted Riemann surface. This was used, for example, in Ref. \cite{CaCaTo09,CaCaTo11,CoTaTo14} to tackle the multi-interval case with PBC.
However, we will not go through this route: in Section \ref{subsec_wave} we rather compute the wavefunction in terms of the correlator of the density field, which should be determined by means of independent methods. Since the theory free, the wavefunction turns out to have a gaussian form: simple enough to be handled in the forthcoming steps. 
We now revert to the problem of computing the reduced density matrices and then the products. The matrix elements of the ground state density matrix in the coherent state basis are simply the product of the wavefunction
\be
\langle \theta |\hat{\rho}|\theta'\rangle=W[\theta]W^*[\theta']\, ,
\ee
where $W^*$ is the complex conjugated.
Considering the system bipartited into two subsystems $A\cup B$, we split the field into two parts $\theta(x)=\theta^A(x)+\theta^B(x)$ with
\be
\theta^A(x)=\begin{cases} \theta(x) \hspace{2pc} & x\in A\\ 0 \hspace{2pc}& x\notin A\end{cases}
\ee
and $\theta^B$ is defined in a similar manner. The reduced density matrix of the subsystem $B$ is obtained tracing away the degrees of freedom in $A$. However, the non-trivial compactification radius plays an essential role: the fields must be identified modulus $2\pi$. Formally we can write
\be
\langle \theta^B|\hat{\rho}^B|(\theta^B)'\rangle=\int \mathcal{D}\theta^A\mathcal{D}(\theta^A)'\, W[\theta^A+\theta^B]W^*[(\theta^A)'+(\theta^B)']\prod_{x\in A}\delta\left(e^{i(\theta^A(x)-(\theta^A)'(x))}\right)\, .
\ee
Furthermore, while computing the products $\hat{\rho}_B=\hat{\rho}_B\times\hat{\rho}_B\times...\times\hat{\rho}_B$ and finally the trace, the bra-ket fields on the subsystem $B$ must be still identified modulus $2\pi$.
The (formal) infinite product of Dirac$-\delta$s is actually less nasty than it could seem at first glimpse: the wavefunction vanishes if the field $\theta$ is not continuous, therefore we are forced to keep $\theta^A(x)-(\theta^A)'(x)$ to be constant in any connected region of the subsystem $A$.
More precisely, imagine $A=\cup_{i=1}^m A_i$ with $A_i$ connected, then the constraint is
\be
\theta^A(x)-(\theta^A)'(x)=2\pi n_i \hspace{2pc} x\in A_i\, .
\ee
Of course, we must sum over all the possible integers $n_i\in \mathbb{Z}$.
Having roughly sketched the battleplan, we now present an helpful short summary of the notation we use, then we proceed with the first step, namely the determination of the ground state wavefunction.

\subsubsection*{List of notations}
First of all, we omit the variable in the derivatives when no ambiguities arise $\partial \theta(x)\equiv \partial_x\theta(x)$: in the next section, where we need to use space and euclidean time, we explicitly write the argument of the derivative. Besides this exception, we stick to this notation.
As we have already done, we consider a bipartition $A\cup B$ and use superscripts $A$ and $B$ for fields living in $A$ and $B$ respectively, being zero otherwise.
Computing the $N^\text{th}$ R\'enyi entropy, we need to introduce a replica space with $N$ copies of the original system.
We use subscripts to distinguish among the copies: given an arbitrary quantity $\rm v$ (it will be either a field, a real number or an integer), with ${\rm v}_j$ we indicate the quantity living in the $j^\text{th}$ copy.
Computing the R\'enyi entropies, different copies will be connected in a cyclic manner, therefore one can identify $j+N\equiv j$.
Furthermore, we call $\tilde{\rm v}_\ell$ the Fourier transform in the replica space, defined according to the following convention
\be\label{eq_fourier_def}
{\rm v}_j=\sum_{\ell=1}^N \frac{e^{i2\pi j \ell/N}}{\sqrt{N}}\tilde{{\rm v}}_\ell\, .
\ee

For the sake of a more compact notation, we introduce a vector-matrix representation for the integrals. For example, given two test functions $\gamma(x)$ and $\beta(x)$ and an operator $\mathcal{A}(x,y)$, we write
\be\label{eq_vec_def}
\gamma^\dagger \beta\equiv\int_0^L \dd x\, \gamma^*(x)\beta(x)\, ,\hspace{2pc}
(\mathcal{A} \beta)(x)\equiv\int_0^L \dd y\, \mathcal{A}(x,y)\beta(y)
\ee
We also make repeated use of the characteristic function $\chi_S(x)$ for an arbitrary set $S$, defined as it follows
\be\label{eq_chi_def}
\chi_S(x)=\begin{cases}1 &\hspace{1pc} x\in S \\ 0 & \hspace{1pc} x \notin S \end{cases}\, .
\ee

\subsection{The ground state wavefunction}
\label{subsec_wave}

As we have already mentioned, the ground state wavefunction can be represented through a path integral in euclidean time \cite{CaCa04,CaCa09}. Even though we ultimately do not explicitly use such a representation, it is nevertheless an useful starting point in studying $W[\theta]$.
Let us consider the euclidean time $\tau$: now the fields live on the euclidean plane $(x,\tau)$. The ground state wavefunction is then the partition function with suitable boundary conditions
\be
W[\theta]\propto\int \mathcal{D}\theta' \mathcal{D}\phi\, \exp\Bigg\{\int_{0}^L\dd x\int_{-\infty}^0\dd \tau\, \frac{i}{\pi}\partial_\tau \theta' \partial_x\phi-\frac{K(x) v(x)}{2\pi}(\partial_x\theta')^2-\frac{v(x)}{2\pi K(x)} (\partial_x \phi)^2\Bigg\}\, .
\ee

Above, the boundary conditions are such that 
\begin{eqnarray}
&&\theta'(x,\tau=0)=\theta(x)\, , \hspace{4pc}\theta'(x,\tau=-\infty)=0
\\
&&\phi(x=0,\tau)=\phi_\text{left}\, , \hspace{4pc} \phi(x=L,\tau)=\phi_\text{right}
\end{eqnarray}
All the remaining boundary conditions are set free. The path integral representation is convenient to deal with the non-trivial boundaries and easily spot a few symmetries. Indeed, in the path integral we can shift the field $\phi$ as per
\be
\phi(x,\tau)=\phi'(x,\tau)+\phi_\text{left}+(\phi_\text{right}-\phi_\text{left})\frac{\int_0^x \dd y\, \frac{K(y)}{v(y)}}{\int_0^L \dd y\, \frac{K(y)}{v(y)}}\, ,
\ee
This implies $\phi'(x=0,\tau)=\phi'(x=L ,\tau)=0$. With this substitution, we get
\be
W[\theta]\propto e^{i\Delta\int_0^L\dd x \theta(x)\frac{K(x)}{v(x)}}\int \mathcal{D}\theta' \mathcal{D}\phi'\, \exp\Bigg\{\int_{0}^L\dd x\int_{-\infty}^0\dd \tau\, \frac{i}{\pi}\partial_\tau \theta' \partial_x\phi'-\frac{ K(x) v(x)}{2\pi}(\partial_x\theta')^2-\frac{v(x)}{2\pi K(x)} (\partial_x \phi')^2\Bigg\}\, ,
\ee
where we set
\be
\Delta=\frac{1}{\pi}\frac{\phi_\text{right}-\phi_\text{left}}{\int_0^L\dd y \frac{K(y)}{v(y)}}\, .
\ee
Above, the path integral describes the ground state wavefunction with trivial Dirichlet boundary conditions in the field $\phi'$, which is fixed to value $0$ at $x=0$ and $x=L$. Rather than aiming for a brute force solution, we can reason as it follows: the path integral is gaussian, therefore the result must be gaussian as well
\begin{multline}
\int \mathcal{D}\theta' \mathcal{D}\phi'\, \exp\Bigg\{\int_{0}^L\dd x\int_{-\infty}^0\dd \tau\, \frac{i}{\pi}\partial_\tau \theta' \partial_x\phi'-\frac{K(x) v(x)}{2\pi}(\partial_x\theta')^2-\frac{v(x)}{2\pi K(x)} (\partial_x \phi')^2\Bigg\}\propto\\
 \exp\Bigg\{-\int\dd x\dd y \,\mathcal{K}(x,y)\theta(x)\theta(y)+\int \dd x\,  C(x)\theta(x)\Bigg\}
\end{multline}
For suitable kernels $\mathcal{K}(x,y)$ and $C(x)$ yet to be determined.
Parity symmetry holds true, i.e. the result does not change if we replace $\theta'(x,\tau)=-\theta'(x,\tau)$ and $\phi'(x,\tau)=-\phi'(x,\tau)$, together with a reflection in the boundary conditions $\theta(x)\to-\theta(x)$. Therefore, we can conclude $C(x)=0$.
Furthermore, the path integral must be real: indeed, taking the complex conjugate is equivalent to setting $\phi'(x,\tau)\to-\phi'(x,\tau)$. This implies that the kernel $\mathcal{K}(x,y)$, which is a symmetric function, must also be real. Based solely on these symmetry-based arguments, we reach the conclusion
\be
W[\theta]\propto \exp\Bigg\{i\Delta\int_0^L\dd x\, \theta(x)\frac{K(x)}{v(x)}-\int\dd x\dd y \,\mathcal{K}(x,y)\theta(x)\theta(y)\Bigg\}\, .
\ee
In principle, the kernel $\mathcal{K}(x,y)$ could be fixed solving the path integral, but we rather prefer to express it in terms of a suitable two-point correlation function, which can be then computed by other means.
From a direct analysis of the gaussian path-integral, we can connect the two-point correlator of $\theta$ with the (inverse of the) kernel $\mathcal{K}$:
\be\label{eq_S_corrtheta}
 \mathcal{K}^{-1}(x,y)=4\Big(\langle \hat{\theta}(x)\hat{\theta}(y)\rangle-\langle \hat{\theta}(x)\rangle\langle\hat{\theta}(y)\rangle\Big)\, .
\ee
However, it is more convenient to look at the conjugate field $\hat{\phi}$.
From the commutation rules, we see that $\partial_x \hat{\phi}$ acts as a functional derivative on the wavefunction
\be
\bra{\theta}\frac{i}{\pi}\partial_x\hat{\phi}\ket{GS}=\frac{\delta}{\delta \theta(x)} W[\theta]\, .
\ee
Which quickly leads to the following simple identity
\be
\mathcal{K}(x,y)=\frac{1}{\pi^2}\partial_x\partial_y \Phi(x,y)\, .
\ee
Above, we defined the connected correlator $\Phi(x,y)=\langle\hat{\phi}(x)\hat{\phi}(y)\rangle-\langle\hat{\phi}(x)\rangle\langle\hat{\phi}(y)\rangle$. 
We stress once again that the connected correlator is independent from the actual boundary conditions $\phi_\text{left}$ and $\phi_\text{right}$. 
Note that, obviously, $\Phi(x=0,y)=\Phi(x=L,y)=0$.
For later use, it is convenient to express the wavefuntion $W[\theta]$ in terms of the derivative of the field $\partial_x\theta$ and of the field at the origin $\theta(0)$. In this case, we get
\be\label{eq_wavefunction}
W[\theta]\propto \exp\Bigg\{\frac{i}{\pi}\theta(0)(\phi_\text{right}-\phi_\text{left})+i\int_0^L\dd x\, \xi(x)\partial_x\theta -\frac{1}{\pi^2}\int\dd x\dd y \,\Phi(x,y)\partial_x\theta(x)\partial_y\theta(y)\Bigg\}\, .
\ee
where we defined
\be
\xi(x)=\Delta\int_x^L\dd y\,  \frac{K(y)}{v(y)}\, .
\ee

\subsection{Warm up: R\'enyi entropy for an interval attached to the boundary}

\begin{figure}[t!]
\includegraphics[width=0.5\textwidth]{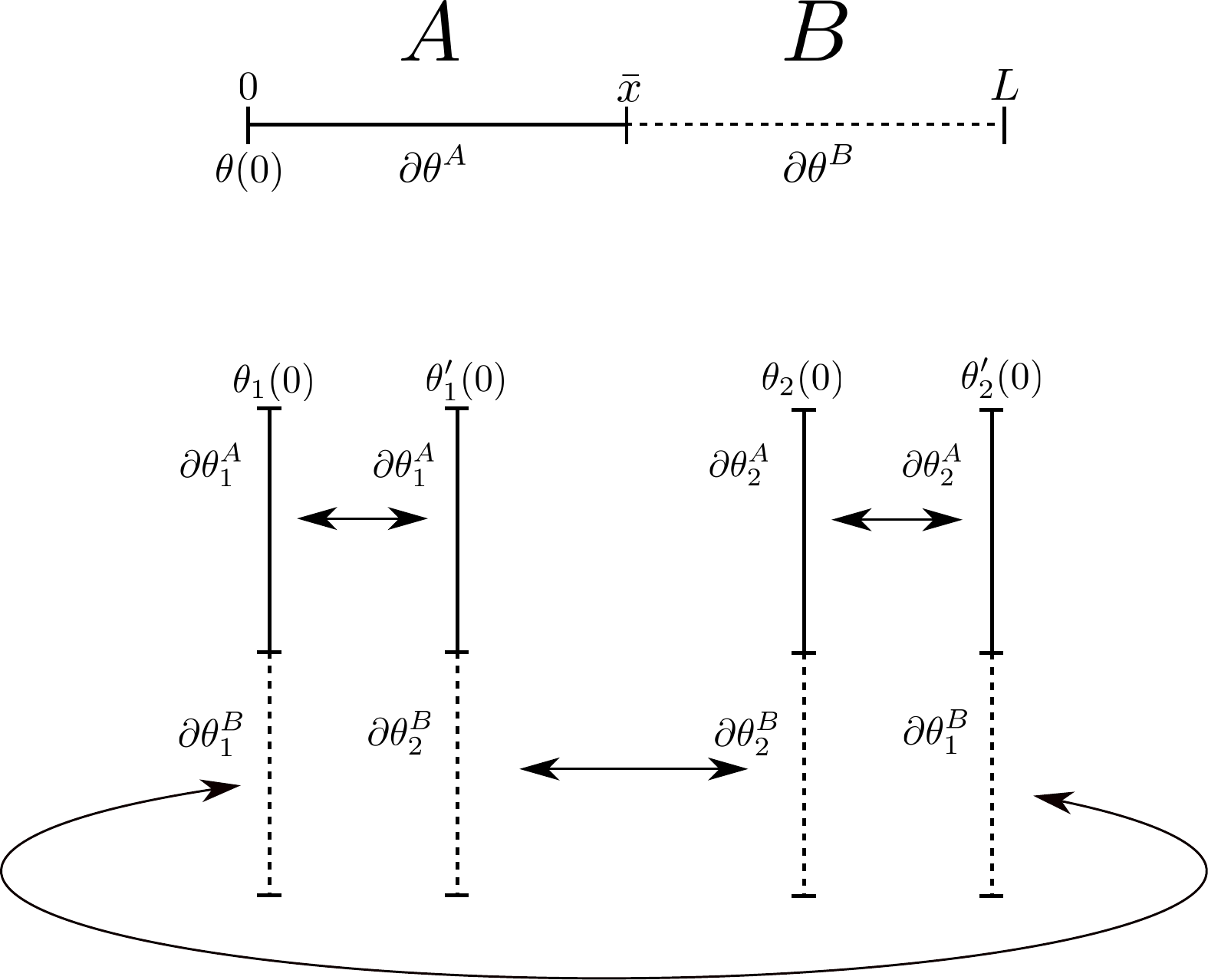}
\caption{\label{fig_2}\emph{Graphical representation of the computation of the trace of powers of the reduced density matrix for a single interval attached to the boundary, according to the notation of Eq. \eqref{eq_rho_tr}. For simplicity, we consider $\text{Tr}\hat{\rho}_B^2$. We represent the real space as a line, the bipartition $A\cup B$ is $A=[0,\bar{x}]$ and $B=[\bar{x},L]$: the part of the system $A$ is represented with a continuous line, while $B$ is associated with a dashed one. The fields $\partial \theta^A$ and $\partial \theta^B$ have as domain $A$ and $B$ respectively. In position $x=0$, there is the additional degree of freedom $\theta(0)$. While computing $\hat{\rho}_B$ the fields in the interval $A$ must be identified modulus a constant offset multiple of $2\pi$, therefore the derivatives of the field are the same (we represent this operation with an arrow). Later, while computing $\hat{\rho}_B^2$ and finally taking the trace, we perform a similar operation identifying the fields $\theta_B$ pairwise.
}
}
\end{figure}

In order to build some insight in the required technical steps, it is useful to start with a simple case, namely we choose  a bipartition $A\cup B$ where $B$ is an interval attached to the boundary $B=[\bar{x},L]$. 
In this case and in the homogeneous setup, the R\'enyi entropies are fixed by the underlying CFT and the Luttinger parameter $K$ does not play any role, except for the Affleck-Ludwig boundary term.
Nevertheless, tackling at first this simpler setup is an useful exercise for the forthcoming, more complicated, computations.
As we anticipated, rather than considering path integrals in the fields $\theta$, it is more convenient to change basis as
\be
\theta(x)=\theta(0)+\int_0^x\dd y\, \partial\theta(y)\, .
\ee
and replace the path integral over the field $\theta$ with an integration over its derivative and the field at the origin $\theta(0)$.
As it is clear from Eq. \eqref{eq_wavefunction}, the wavefunction can be factorized into a term which depends only on $\theta(0)$ and another one which accounts for the derivatives of the field.
For the sake of simplicity, we introduce the following notation
\be\label{eq_W}
W(\theta)=e^{\frac{i}{\pi}\theta(0)(\phi_\text{right}-\phi_\text{left})} \mathcal{W}\begin{pmatrix} \partial\theta^A\\ \partial\theta^B \end{pmatrix}\, .
\ee
While gluing the fields together modulus $2\pi$, the derivatives coincide. 
The identification of the fields can be understood with the help of Fig. \ref{fig_2}: the difference of two fields connected by two arrows is an integer multiple of $2\pi$. This can be implemented identifying the field derivatives and adding the proper constraints at the edges of the intervals, namely (see also Fig. \ref{fig_2})
\be
\theta^j(0)-(\theta^j)'(0)=2\pi n_j\, ,
\hspace{3pc}
\theta_j(0)-\theta_{j+1}(0)+\chi^\dagger_{A} [\partial\theta^A_{j}-\partial\theta^A_{j+1}]=2\pi p_j\, ,
\ee
with $n_j,p_j\in \mathbb{Z}$. Above, we used the vector notation Eq. \eqref{eq_vec_def} and the characteristic function defined in Eq. \eqref{eq_chi_def}.

We can conveniently subtract the first constraint from the second and redefine new integers $m_j=p_j-n_j$, then write the $N^\text{th}$ power of the reduced density matrix as it follows
\begin{multline}\label{eq_rho_tr}
\text{Tr}\hat{\rho}_B^N=\frac{1}{\mathcal{Z}^N}\sum_{\{n_j\}_{j=1}^N,\, \{m_j\}_{j=1}^{N-1}}\int \dd^N \theta(0) \mathcal{D}^N\partial\theta^A \mathcal{D}^N\partial\theta^B  e^{-i2 \sum_{j=1}^N n_j(\phi_\text{right}-\phi_\text{left})}\prod_{j=1}^N\mathcal{W}\begin{pmatrix} \partial\theta_j^A\\ \partial\theta_j^B \end{pmatrix} \mathcal{W}^*\begin{pmatrix} \partial\theta_j^A\\ \partial\theta_{j+1}^B \end{pmatrix}\times \\
\prod_{j=1}^{N-1}\delta\left(\theta_j(0)-\theta_{j+1}(0)+\chi_A^\dagger[ \partial\theta_j^A-\partial\theta^A_{j+1}]-2\pi m_j\right)\, .
\end{multline}
PBC on the replica space are imposed, in such a way that $\partial\theta^B_{N+1}=\partial\theta^B_1$. The partition function $\mathcal{Z}$ must be fixed imposing $\text{Tr}\hat{\rho}^B=1$, as we will do later on.
The summation is over all the possible integers $\{n_j\}_{j=1}^N,\, \{m_j\}_{j=1}^{N-1}$. We stress that the index of the integers $\{m_j\}_{j=1}^{N-1}$ runs from $j=1$ to $j=N-1$: indeed, the $N^\text{th}$ constraint can be obtained as a linear combination of the others.
First of all, note that the sum over the integers $n_j$ imposes
\be\label{eq_left_right}
\phi_\text{right}-\phi_\text{left}\in \pi\mathbb{Z}\, ,
\ee
otherwise the result vanishes. As we anticipated, this quantization of the boundary conditions is required by the presence of the $U(1)$ conserved charge in the microscopic model \cite{Ca04}.
Furthermore, the absence of any $\theta_j(0)$ dependence in the wavefunction $\mathcal{W}$ allows us to use the integration over $\theta_j(0)$ to get rid of the $\delta$ constraints. 
Having $N-1$ constraints and $N$ integration variables $\theta_j(0)$, we are still left with a variable to be integrated, which we choose to be $\theta_1(0)$. Furthermore, we have $N-1$ summations over the integers $m_j$ and $N$ summations over $n$.
We are left with the following formal expression
\be
\text{Tr}\hat{\rho}_B^N=\frac{1}{\mathcal{Z}^N}\left(\int \dd \theta_1(0)\, 1\right)\left(\sum_m 1 \right)^{N-1}\left(\sum_n 1\right)^N\int  \mathcal{D}^N\partial\theta^A \mathcal{D}^N\partial\theta^B \prod_{j=1}^N\mathcal{W}\begin{pmatrix} \partial\theta_j^A\\ \partial\theta_j^B \end{pmatrix} \mathcal{W}^*\begin{pmatrix} \partial\theta_j^A\\ \partial\theta_{j+1}^B \end{pmatrix}\, ,
\ee
The integrals and summations above are on the constant function $1$ and are formally divergent quantities.
We are not entitled to discard the divergent prefactor, which we will analyze later on.
The path integral is now a simple gaussian integration. Note that, because of the PBC on the replica space, the linear part in the exponential of $\mathcal{W}$ gets averaged to zero and we are left with the following expression
\be
\int  \mathcal{D}^N\partial\theta^A \mathcal{D}^N\partial\theta^B 
\exp\Bigg\{-\frac{1}{\pi^2}\sum_{j=1}^N  (\partial\theta^A_j+\partial\theta^B_j)^\dagger \Phi(\partial\theta^A_j+\partial\theta^B_j)+(\partial\theta^A_j+\partial\theta^B_{j+1})^\dagger\Phi(\partial\theta^A_j+\partial\theta^B_{j+1}) \Bigg\}\, .
\ee

Above, we use the vector-matrix notation \eqref{eq_vec_def} for the integrals.
The quadratic term in the exponential acquires a simple block-diagonal form if we consider the Fourier transform in the replica space. We use the convention of Eq. \eqref{eq_fourier_def} and obtain the following expression
\be\label{eq_int_gauss}
\int  \mathcal{D}^N\partial\theta^A \mathcal{D}^N\partial\theta^B \exp\Bigg\{-\frac{1}{\pi^2}\sum_{\ell=1}^N \partial\tilde{\theta}_\ell^\dagger (\Phi+\Phi_{\ell/N})\partial\tilde{\theta}_\ell \Bigg\}=\sqrt{\frac{1}{\det\left(\Phi/\pi^3+\Phi_{\ell/N}/\pi^3\right)}}
\ee
Above, we defined
$\partial\tilde{\theta}_\ell=\partial\tilde{\theta}^A_\ell+\partial\tilde{\theta}^B_\ell$
and $\Phi_\Omega$ is defined as per Eq. \eqref{eq_phiomega}.
Putting together all the terms we get
\be\label{eq_tr_B_N}
\text{Tr}\hat{\rho}_B^N= \frac{1}{\mathcal{Z}^N} \left(\int \dd \theta_1(0)\, 1\right)\left(\sum_m 1 \right)^{N-1}\left(\sum_n 1\right)^N\prod_{\ell=1}^N\sqrt{\frac{1}{\det\left(\Phi/\pi^3+\Phi_{\ell/N}/\pi^3\right)}}
\ee
where $\det\left(\Phi/\pi^3+\Phi_{\ell/N}/\pi^3\right)$ must be understood as the determinant of the operator $\Phi(x,y)/\pi^3+\Phi_{\ell/N}(x,y)/\pi^3$.
Imposing $\text{Tr}\hat{\rho}_B=1$ we are forced to require
\be\label{eq_normalization}
\mathcal{Z}=\left(\int \dd \theta_1(0)\, 1\right)\left(\sum_n 1\right)\sqrt{\frac{1}{\det\left(2\Phi/\pi^3\right)}}
\ee
Employing this result in Eq. \eqref{eq_tr_B_N}, we arrive at the following expression for the R\'enyi entropy
\be
\mathcal{S}^N_{[\bar{x},L]}=\frac{1}{1-N}\log\left[\prod_{\ell=1}^{N-1}\sqrt{\frac{1}{\det\left(\frac{1+\Phi^{-1}\Phi_{\ell/N}}{2}\right)}}\right]+\log\left[\frac{\int \dd \theta_1(0)\, 1}{\sum_m 1}\right]\, .
\ee

This concludes the derivation for the interval attached to the boundary.
Above, we can recognize two terms: the first one actually depends on the chosen interval and, in the homogeneous case, it contains the CFT scaling. It is easy to check that, in the homogeneous case, this term is completely independent of the choice of the Luttinger parameter $K$. Even though an analytical extraction of the CFT result from the above expression is a hard task, a simple numerical test with the discretization presented in Appendix \ref{app_latt} ensures the matching with the CFT scaling.
The second term does not depend on the actual interval neither on the index of the R\'enyi entropy and it is undetermined within our approach: we interpret it as the ground state degeneracy resulting in the Affleck-Ludwig boundary entropy \cite{AfLu91,ZhBaFj06}
\be\label{eq_g_factor}
g=\frac{\int \dd \theta_1(0)\, 1}{\sum_m 1}\, .
\ee
Indeed, the very same term appears also in computing other bipartitions, strengthening once again the above identification: we will see in the next section in the case of an interval in the middle of the system and then later for general bipartitions in Section \ref{sec_gen_bip}. 
A first principle derivation of the ground state degeneracy $g$ within the present method would require a proper regularization of the integral (and sum) over the target space appearing into Eq. \eqref{eq_g_factor}. Such a regularization lays beyond our current understanding.

\subsection{An interval in the middle}
\label{subsec_Renyi_inter}

\begin{figure}[t!]
\includegraphics[width=0.5\textwidth]{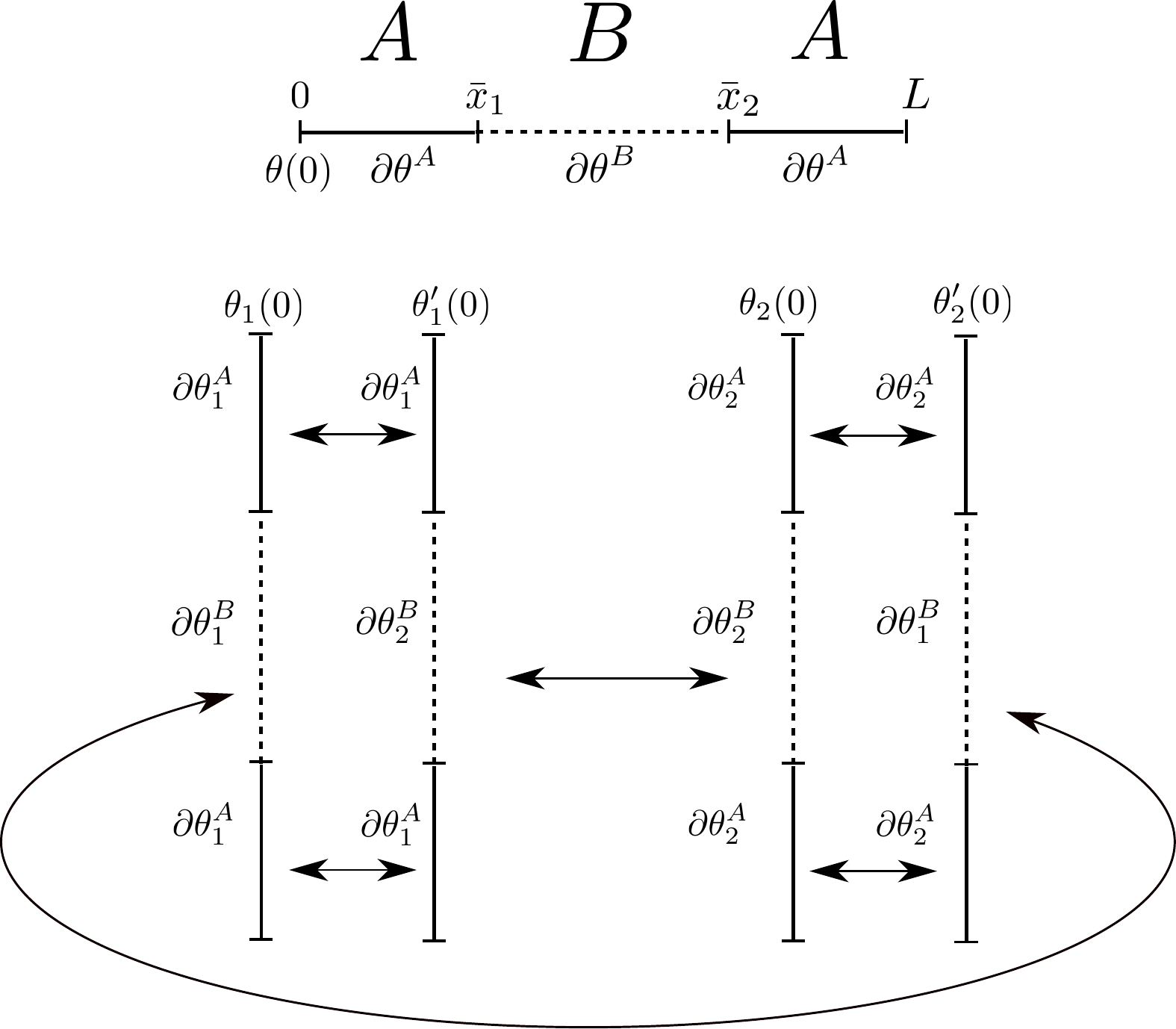}
\caption{\label{fig_3}\emph{Graphical representation of the trace of powers of the reduced density matrix with a bipartition $A\cup B$ where $B$ is chosen as an interval in the middle of the system. The case $\text{Tr}\hat{\rho}_B^2$ is depicted, the same notation of Fig. \ref{fig_2} is used.}}
\end{figure}
We now turn to the main case of interest, namely $B=[x_1,x_2]$ with $0<x_1<x_2<L$. We still conveniently integrate over the derivative of the field and its value at the boundary, keeping the same notation as Eq. \eqref{eq_W}.
The trace of the reduced density matrix has a similar expression to Eq. \eqref{eq_rho_tr}, however with a further delta function (see Fig. \ref{fig_3}).
Indeed, the following constraints must be enforced
\be
\theta_j(0)-\theta_j'(0)=2\pi n_j\, ,
\ee
\be
\theta_j'(0)-\theta_{j+1}(0)+\chi_{[0,x_1]}^\dagger [\partial\theta^A_{j}-\partial\theta^A_{j+1}]=2\pi p_j\, ,
\ee
\be
\theta_j(0)-\theta_j'(0)+\chi_{[x_1,x_2]}^\dagger[\partial\theta^B_j-\partial\theta^B_{j+1}]=2\pi q_j\, .
\ee
It is convenient to subtract from the second and third constraint the first one, in this way we reach the following expression for the trace of the reduced density matrix
\begin{multline}\label{eq_traceB_N}
\text{Tr}\hat{\rho}_B^N=\frac{1}{\mathcal{Z}}\sum_{\{n_j\}_{j=1}^N,\, \{m_j\}_{j=1}^{N-1},\{z_j\}_{j=1}^{N-1}}\int \dd^N \theta(0) \mathcal{D}^N\partial\theta^A \mathcal{D}^N\partial\theta^B  e^{-i2 \sum_{j=1}^N n_j(\phi_\text{right}-\phi_\text{left})}\prod_{j=1}^N\mathcal{W}\begin{pmatrix} \partial\theta_j^A\\ \partial\theta_j^B \end{pmatrix} \mathcal{W}^*\begin{pmatrix} \partial\theta_j^A\\ \partial\theta_{j+1}^B \end{pmatrix}\times \\
\prod_{j=1}^{N-1}\delta\left(\theta_j(0)-\theta_{j+1}(0)+\chi^\dagger_{[0,x_1]}[ \partial\theta^A_j-\partial\theta^A_{j+1}]-2\pi m_j\right)\prod_{j=1}^{N-1}\delta\left(\chi_{[x_1,x_2]}^\dagger[\partial\theta^B_j-\partial\theta^B_{j+1}]-2\pi z_j\right)\, .
\end{multline}
As we did in the previous section, we use the integration over $\theta_j(0)$ to get rid of the first set of constraints. As before, the summation over $n_j$ becomes trivial once Eq. \eqref{eq_left_right} is imposed.
However, after these operations, a set of non trivial constraints is still left out
\begin{multline}
\text{Tr}\hat{\rho}_B^N=\frac{\left(\int \dd \theta_1(0)\, 1\right)\left(\sum_m 1 \right)^{N-1}\left(\sum_n 1\right)^N}{\mathcal{Z}^N}\times\\
\sum_{\{z_j\}_{j=1}^{N-1}}\int \mathcal{D}^N\partial\theta^A \mathcal{D}^N\partial\theta^B  \prod_{j=1}^N\mathcal{W}\begin{pmatrix} \partial\theta_j^A\\ \partial\theta_j^B \end{pmatrix} \mathcal{W}^*\begin{pmatrix} \partial\theta_j^A\\ \partial\theta_{j+1}^B \end{pmatrix}
\prod_{j=1}^{N-1}\delta\left(\chi_{[x_1,x_2]}^\dagger[\partial\theta^B_j-\partial\theta^B_{j+1}]-2\pi z_j\right)\, .
\end{multline}
Above, note the appearance of the same prefactor that, in the previous section, resulted in the ground state degeneracy \eqref{eq_g_factor}. The normalization $\mathcal{Z}$ is of course independent on the chosen bipartition and we already computed it in Eq. \eqref{eq_normalization}. Employing this result together with Eq. \eqref{eq_g_factor} we can write
\be
\frac{\left(\int \dd \theta_1(0)\, 1\right)\left(\sum_m 1 \right)^{N-1}\left(\sum_n 1\right)^N}{\mathcal{Z}^N}=g^{1-N}\left[\det\left(2\Phi/\pi^3\right)\right]^{N/2}\, .
\ee
In order to take care of the $\delta-$functions, we use a proper integral representation
\be
\delta(x)=\int \frac{\dd\lambda}{2\pi} e^{i\lambda x}\,
\ee
for each constraint. Therefore
\begin{multline}\label{eq_calculation_rho}
\text{Tr}\hat{\rho}_B^N=g^{1-N}\left[\det\left(2\Phi/\pi^3\right)\right]^{N/2}\sum_{\{z_j\}_{j=1}^{N-1}}\int \frac{\dd^N \lambda}{(2\pi)^N}2\pi\delta(\lambda_N)\int \mathcal{D}^N\partial\theta^A \mathcal{D}^N\partial\theta^B  \prod_{j=1}^N\mathcal{W}\begin{pmatrix} \partial\theta_j^A\\ \partial\theta_j^B \end{pmatrix} \mathcal{W}^*\begin{pmatrix} \partial\theta_j^A\\ \partial\theta_{j+1}^B \end{pmatrix}\times \\
\exp\Bigg\{i\sum_{j=1}^N\lambda_j\left(\chi_{[x_1,x_2]}[\partial\theta^B_j-\partial\theta^B_{j+1}]-2\pi z_j\right)\Bigg\}\,.
\end{multline}
Above, we conveniently added a dummy integration over an additional field $\lambda_N$ that is then fixed at value $0$ by the Dirac$-\delta$. This makes the integrand more symmetric and easier to be handle. Furthermore, we insert another integral representation for $\delta(\lambda_N)$ 
\be
2\pi\delta(\lambda_N)=\int \dd \omega\, e^{-i\omega \lambda_N}\, .
\ee
Rather than using $\lambda_j$, we introduce new variables $\zeta_j$ in such a way
\be
\lambda_j=\sum_{i=1}^j \zeta_i\, .
\ee
Which implies $\lambda_{j+1}-\lambda_{j}=\zeta_j$. Obviously, we have $\lambda_N=\sum_{i=1}^N \zeta_i$.
The real numbers $\zeta_j$ and integers $z_j$ are Fourier-transformed in the replica space according to the usual convention Eq. \eqref{eq_fourier_def} and we reach the following expression
\begin{multline}\label{eq_tr_68}
\text{Tr}\hat{\rho}_B^N=g^{1-N}\left[\det\left(2\Phi/\pi^3\right)\right]^{N/2}\sum_{\{z_j\}_{j=1}^{N-1}}\int \dd \omega\int \frac{\dd^N \zeta}{(2\pi)^N}e^{-i\sum_{j=1}^N(\omega+2\pi \sum_{i=j}^Nz_i) \zeta_j}  \times \\
\int \mathcal{D}^N\partial\theta^A \mathcal{D}^N\partial\theta^B\exp\Bigg\{\sum_{\ell=1}^N \Bigg[-\frac{1}{\pi^2}\partial\tilde{\theta}_\ell^\dagger (\Phi+\Phi_{\ell/N})\partial\tilde{\theta}_\ell+
\frac{i}{2}[\tilde{\zeta}_\ell]^* \chi_{[x_1,x_2]}^\dagger\partial\tilde{\theta}_\ell+ \frac{i}{2}\tilde{\zeta}_\ell\partial\tilde{\theta}_\ell^\dagger\chi_{[x_1,x_2]}\Bigg]\Bigg\}\, .
\end{multline}

We now change variable in the summation and define new integers
$m_j=\sum_{i=j}^N z_i$.
This is a one-to-one map provided we set $m_N=0$. We use these new integers to perform the summation, since the constraint associated to the $\zeta$ variables gets simplified.
We now compute the gaussian path integral, leaving out the integration over $\lambda_j$ and $\omega$ for a later time.
We get rid of the imaginary part performing an analytic continuation to complex fields
\be
\partial \theta^A_j\to \partial \theta^A_j+i\delta \partial\theta^A_j\, ,\hspace{2pc}\partial \theta^B_j\to \partial \theta^B_j+i\delta \partial\theta^B_j
\ee
In terms of the fields in the Fourier space, this amounts to a transformation
\be
\partial\tilde{\theta}_\ell\to \partial\tilde{\theta}_\ell+i\delta\partial\tilde{\theta}_\ell\,,\hspace{2pc}
\partial\tilde{\theta}_\ell^\dagger\to \partial\tilde{\theta}_\ell^\dagger+i\delta\partial\tilde{\theta}_\ell^\dagger
\ee
The shift must be chosen in such a way as to absorb the linear term, i.e. we require
\be
\delta \partial\tilde{\theta}_\ell=\frac{\pi^2}{2}\tilde{\zeta}^\ell(\Phi+\Phi_{\ell/N})^{-1}\chi_{[x_1,x_2]}\, .
\ee
After such a shift, we get
\begin{multline}
\int \mathcal{D}^N\partial\theta^A \mathcal{D}^N\partial\theta^B\exp\Bigg\{\sum_{\ell=1}^N \Bigg[-\frac{1}{\pi^2}\partial\tilde{\theta}_\ell^\dagger (\Phi+\Phi_{\ell/N})\partial\tilde{\theta}_\ell+
\frac{i}{2}[\tilde{\zeta}_\ell]^* \chi_{[x_1,x_2]}^\dagger\partial\tilde{\theta}_\ell+ \frac{i}{2}\tilde{\zeta}_\ell\partial\tilde{\theta}_\ell^\dagger\chi_{[x_1,x_2]}\Bigg]\Bigg\}=\\
\exp\Bigg\{-\frac{\pi^2}{2}\sum_{\ell=1}^N|\tilde{\zeta}_\ell|^2\chi_{[x_1,x_2]}^\dagger(\Phi+\Phi_{\ell/N})^{-1}\chi_{[x_1,x_2]}\Bigg\}\int \mathcal{D}^N\partial\theta^A \mathcal{D}^N\partial\theta^B\exp\Bigg\{\sum_{\ell=1}^N \Bigg[-\frac{1}{\pi^2}\partial\tilde{\theta}_\ell^\dagger (\Phi+\Phi_{\ell/N})\partial\tilde{\theta}_\ell\Bigg]\Bigg\}\, .
\end{multline}
The last path integral is now performed as we did in the previous section. We then reach the following expression
\begin{multline}\label{eq_rhon_74}
\text{Tr}\hat{\rho}_B^N=g^{1-N}\prod_{\ell=1}^{N-1}\sqrt{\frac{1}{\det\left(\frac{1+\Phi^{-1}\Phi_{\ell/N}}{2}\right)}}\times\\
 \sum_{\{m_j\}_{j=1}^{N-1}}
\int \dd \omega\int \frac{\dd^N\zeta}{(2\pi)^N}   \exp\Bigg\{\sum_{\ell=1}^N\Bigg[-\frac{\pi^2}{4}|\tilde{\zeta}_\ell|^2\mathcal{I}_{\ell/N}-\frac{i}{2}(\omega\delta_{\ell,N}+2\pi [\tilde{m}_\ell]^*) \tilde{\zeta}_\ell-\frac{i}{2}(\omega\delta_{\ell,N}+2\pi \tilde{m}_\ell) [\tilde{\zeta}_\ell]^*\Bigg]\Bigg\}\, .
\end{multline}
Where we defined the real numbers $\mathcal{I}_{\ell/N}$ as
\be
\mathcal{I}_{\ell/N}=\chi_{[x_1,x_2]}^\dagger(\Phi+\Phi_{\ell/N})^{-1}\chi_{[x_1,x_2]}\, .
\ee
We stress that this definition is equivalent to Eq. \eqref{eq_Il_def}.
The last step requires computing the gaussian integral in the variables $\zeta$ and than that on $\omega$, which leads to the following result 
\be
\text{Tr}\hat{\rho}_B^N=g^{1-N}\frac{1}{(2\pi)^{N-1}}\prod_{\ell=1}^{N-1}\sqrt{\frac{1}{\det\left(\frac{1+\Phi^{-1}\Phi_{\ell/N}}{2}\right)}}\prod_{\ell=1}^{N-1}\sqrt{\frac{4}{\pi\mathcal{I}_{\ell/N}}}
\sum_{\{m_j\}_{j=1}^{N-1}}  \exp\Bigg\{-\sum_{\ell=1}^{N-1}\frac{4}{\mathcal{I}_{\ell/N}}|\tilde{m}_{\ell}|^2\Bigg\}\, .
\ee
Note that $\text{Tr}\hat{\rho}_B=1$, as it should be.
Since the sum must be performed on real integers $m_j$, it is more convenient to write the exponent in the real space rather than in the Fourier one. Therefore, let us define the matrix $\mathcal{M}_{ab}$ as per Eq. \eqref{eq_M_def} and write
\be
\text{Tr}\hat{\rho}_B^N=g^{1-N}\prod_{\ell=1}^{N-1}\sqrt{\frac{1}{\pi^3\mathcal{I}_{\ell/N}}\frac{1}{\det\left(\frac{1+\Phi^{-1}\Phi_{\ell/N}}{2}\right)}}
\sum_{\{m_j\}_{j=1}^{N-1}}  \exp\Bigg\{-4\sum_{a,b=1}^{N-1}\mathcal{M}_{ab}m_a m_b\Bigg\}\, .
\ee
Taking the logarithm, we are immediately lead to Eq. \eqref{eq_result}.

\subsection{Generalization to arbitrary bipartitions}
\label{sec_gen_bip}

The same analysis of the previous section can be further generalized to arbitrary bipartitions: the main difficulty consists in choosing the right notation.
We now imagine a collection of points on the real axis $\{x_i\}_{i=1}^{N_{I}-1}$ such that $0<x_i<x_{i+1}<L$ and consider the bipartition
\be
A=\cup_{i=0}^{\lfloor\frac{N_I-1}{2}\rfloor}[x_{2i},x_{2i+1}]\,, \hspace{4pc} B=\cup_{i=1}^{\lceil \frac{N_I-1}{2}\rceil} [x_{2i-1},x_{2i}]\, .
\ee
Above, we set $x_0=0$ and $x_{N_I}=L$.
We now focus on the $N^\text{th}$ R\'enyi. We keep the same notation for the fields as in the previous section and retain the vector-matrix notation as well. We define $\partial\theta_j=\partial\theta^A_j+\partial\theta^B_j$. Then, it is not hard to understand that the non trivial compactification radius induces the following set of non trivial constraints
\be
\chi_{[x_i,x_{i+1}]}^\dagger(\partial\theta_j-\partial\theta_{j+1})=2\pi z_j^i \,, \hspace{2pc} 1\le i \le N_I-2\, ,\hspace{1pc} 1\le j\le N-1\, ,\hspace{1pc}z^i_j \in \mathbb{Z}
\ee
If we choose $N_I=3$ we are back to the case of a single interval in the middle studied in the previous section. We then need to compute the following path integral
\begin{multline}
\text{Tr}\hat{\rho}_B^N=g^{1-N}\left[\det\left(2\Phi/\pi^3\right)\right]^{N/2}\times\\\sum_{\{z_j^i\}}\int \mathcal{D}^N\partial\theta^A \mathcal{D}^N\partial\theta^B  \prod_{j=1}^N\mathcal{W}\begin{pmatrix} \partial\theta_j^A\\ \partial\theta_j^B \end{pmatrix} \mathcal{W}^*\begin{pmatrix} \partial\theta_j^A\\ \partial\theta_{j+1}^B \end{pmatrix}
\prod_{j=1}^{N-1}\prod_{i=1}^{N_I-2}\delta\left(\chi_{[x_i,x_{i+1}]}^\dagger(\partial\theta_j-\partial\theta_{j+1})-2\pi z_j^i \right)\, .
\end{multline}
Above, the summation is over all the possible integers $z_j^i$.
As before, we introduce an integral representation of the $\delta$ function for each constraint
\begin{multline}
\text{Tr}\hat{\rho}_B^N=g^{1-N}\left[\det\left(2\Phi/\pi^3\right)\right]^{N/2}\sum_{\{z_j^i\}}\int \frac{\dd^{N(N_I-2)}\lambda}{(2\pi)^{N(N_I-2)}}\prod_{i=1}^{N_I-2}\Big[2\pi\delta(\lambda^{i}_N)\Big]e^{-2\pi i\sum_{i,j} \lambda^{i}_j z^i_j }\times\\
\int \mathcal{D}^N\partial\theta_A \mathcal{D}^N\partial\theta_B  \prod_{j=1}^N\mathcal{W}\begin{pmatrix} \partial\theta_j^A\\ \partial\theta_j^B \end{pmatrix} \mathcal{W}^*\begin{pmatrix} \partial\theta_j^A\\ \partial\theta_{j+1}^B \end{pmatrix}
\exp\left[i\sum_{i,j}\lambda^i_j\chi_{[x_i,x_{i+1}]}^\dagger(\partial\theta_j-\partial\theta_{j+1}) \right]\, .
\end{multline}
Now, generalizing what we have previously done, we change variables for  $\lambda^i_j$ and on the integers $z^i_j$ as
\be
\lambda_j^i=\sum_{j'=1}^j \zeta_{j'}^i\, \hspace{3pc}
m_{j}^i=\sum_{j'=j}^N z^i_{j'}\, .
\ee
Rather than summing over $z^i_j$ we can sum over the integers $m^i_j$. We furthermore introduce additional variables $\omega^i$ to write an integral representation of $\delta(\lambda^i_N)$. Once we consider the Fourier transform in the replica space, we reach a generalization of Eq. \eqref{eq_tr_68}

\begin{multline}
\text{Tr}\hat{\rho}_B^N=g^{1-N}\left[\det\left(2\Phi/\pi^3\right)\right]^{N/2}\sum_{\{m_i^j\}}\int \dd^{N_I-2} \omega\int \frac{\dd^{N(N_I-2)} \zeta}{(2\pi)^{N(N_I-2)}}e^{-i\sum_{i=1}^{N_I-2}\sum_{j=1}^N(\omega^i+2\pi m^i_j) \zeta^i_j}  \times \\
\int \mathcal{D}^N\partial\theta^A \mathcal{D}^N\partial\theta^B\exp\Bigg\{\sum_{\ell=1}^N \Bigg[-\frac{1}{\pi^2}\partial\tilde{\theta}_\ell^\dagger (\Phi+\Phi_{\ell/N})\partial\tilde{\theta}_\ell+
\frac{i}{2}\sum_{i=1}^{N_I-2}[\tilde{\zeta}^i_\ell]^* \chi_{[x_i,x_{i+1}]}^\dagger\partial\tilde{\theta}_\ell+ \frac{i}{2}\sum_{i=1}^{N_I-2}\tilde{\zeta}^i_\ell\partial\tilde{\theta}_\ell^\dagger\chi_{[x_{i},x_{i+1}]}\Bigg]\Bigg\}\, .
\end{multline}

Above, $\Phi_\Omega$ is still defined as Eq. \eqref{eq_phiomega}, with the difference that now $B$ is made of several disconnected intervals.

The gaussian path integral is carried out through an analytical continuation in the complex plane, as we did in the single interval case and we then reach the analogue of Eq. \eqref{eq_rhon_74}, but with extra indices. After that, the gaussian integral on the $\zeta$ and $\omega$ variables can be computed and the following final expression is reached 

\be\label{sn_general_int}
\mathcal{S}^N=\frac{1}{1-N}\log\left[\prod_{\ell=1}^{N-1}\sqrt{\frac{1}{\det(\pi^3\mathcal{I}_{\ell/N})\det\left(\frac{1+\Phi^{-1}\Phi_{\ell/N}}{2}\right)}}\sum_{\{m^i_j\}}
 \exp\Bigg\{-4\sum_{a,b}^{N-1}\sum_{i,i'}^{N_I-2}\mathcal{M}_{ab}^{ii'}m^i_a m^{i'}_b\Bigg\}\right]+\log g\, .
\ee

Above, we defined the tensor
\be
\mathcal{M}_{ab}^{jj'}=\sum_{\ell=1}^{N-1} \frac{e^{-i2\pi \ell(a-b)/N}}{N}[\mathcal{I}_{\ell/N}^{-1}]^{jj'}\, ,\hspace{2pc} \mathcal{I}_{\ell/N}^{ii'}=\chi_{[x_i,x_{i+1}]}^\dagger(\Phi+\Phi_{\ell/N})^{-1}\chi_{[x_{i'},x_{i'+1}]}\,.
\ee

It is worth mentioning that, within the homogeneous case, we could generalize the content of Section \ref{sec_hom_ren}: conformal symmetry fixes the R\'enyi entropies apart from an additional contribution $(1-N)^{-1}\log\mathcal{F}_N$. $\mathcal{F}_N$ is a function of all the four-point conformal invariant ratios which can be constructed with the extrema of the intervals (and the boundary points). The R\'enyi entropies at the free fermion point are known explicitly for arbitrary bipartitions (see e.g. Ref. \cite{CaFoHu05}), thus by comparison with Eq. \eqref{sn_general_int} it is possible to fix $\mathcal{F}_N$ in terms of Riemann Theta functions, generalizing Eq. \eqref{eq_FN}.

\section{Conclusions and outlook}
\label{sec_conclusion}

This work has been dedicated to the computation of the R\'enyi entanglement entropies of integer index for the 1d compactified massless boson. While this problem has already been posed (and solved \cite{CaCaTo09,CaCaTo11,CoTaTo14}) for periodic boundary conditions, here we rather address open boundary conditions, which are of utmost importance for tensor networks simulations and the low energy physics of confined systems \cite{DuStViCa17}.
We address this problem for the ground state of an inhomogeneous generalization of the compactified boson, described by locally-varying sound velocity and Luttinger parameter: in this case, we provide Fredholm determinant-like expressions which generalize previously known methods for free systems \cite{CaHu09}.
Within the homogeneous case, which is described by conformal symmetry, we pushed further our analytical investigation reducing the problem to solving linear-integral equations and computing Riemann Theta functions, which account for the role of the non-trivial compactification radius.
Several open questions and interesting applications are left for future investigations.
We presented a non-trivial benchmark of our result on free fermions systems where the analytical expression was known, but the final goal is the description of ground states of interacting models with $K\ne 1$: it would be extremely interesting to compare our findings with DMRG simulations.
We provided general expressions for the R\'enyi entropies having as an input the density-density correlation function: its application to spatially inhomogeneous or out-of-equilibrium setups is a natural question to be investigated.
For what it concerns possible open issues, there are at least three main points.

The first question concerns the determination of the Affleck-Ludwig boundary term within our approach, which leaves it undetermined in absence of a proper not-yet-devised regularization scheme. Within the homogeneous case, we can match this constant with the existing literature, but the application of our findings to inhomogeneous systems ultimately needs a direct computation of the ground state degeneracy.

Secondarily, even considering the homogeneous case, the computation of the R\'enyi entropies requires solving some linear integral equations, task that we performed numerically: having a fully analytical expression would be extremely important, especially for a systematic small and large distance expansion.

The third natural question concerns the possibility of analytically continuing our result to real R\'enyi indices, finally reaching the Von Neumann entanglement entropy. Solving this surely appealing problem seems quite hard, due to the intricate index-dependence of our result: a similar question has already been asked in the closely related PBC problem of disjoint intervals \cite{CaCaTo09,CaCaTo11,CoTaTo14} but, as far as we know, it does not have a direct solution yet, albeit numerical approaches can be used \cite{AgHeJaKa14,DeCoTo15}.
In Ref. \cite{RaGl12,RuToCa18} the PBC problem has been attacked with conformal blocks techniques, resulting in an expansion of the R\'enyi entropies in the distance among the intervals. Remarkably, each term of the series can be analytically continued and an expansion for the Von Neumann entropy is achieved. However, establishing the connection with the exact result of Ref. \cite{CaCaTo09,CaCaTo11,CoTaTo14} is a difficult question. It is natural to wonder if conformal blocks techniques can be used to study the OBC case as well.
Lastly, we would like to mention the intriguing possibility of extending the methods here presented to other measures of entanglement, for example the entanglement negativity \cite{Pe96,ZyHo98,Zy99,LeeKim00,EiPl99,ViWe02,Pl05},  using the replica approach in the path integral formalism \cite{CaCaTo12,CaCaTo13}.

\section*{Acknowledgements}

I am grateful to J. Dubail and J.-M. St\'ephan for joint collaboration on closely related topics. I am also indebted to them, together with P. Calabrese, for having encouraged me in completing this project and for their extremely valuable comments on the manuscript, for which I thank P. Ruggiero as well.
I acknowledge the support from the European
Research Council under ERC Advanced grant
743032 DYNAMINT.

\appendix

\section{R\'enyi entropies and lattice discretization}
\label{app_latt}

In this short appendix we briefly discuss the numerical computation of Eq. \eqref{eq_result}. We consider the homogeneous case where the correlation functions of the density field are analytically known: addressing inhomogeneous setups requires as a further ingredient the numerical computation of the correlations.
Without loss of generality, we consider the system having unitary length, thus $L=1$ and the continuous system lives on a finite interval $[0,1]$. This interval is then discretized into $N_d$ points, being then the lattice spacing $a=1/N_d$.
In the homogeneous case \eqref{eq_corr_hom}, correlation functions are $\Phi(x,y)= K\times[f(x-y)-f(x+y)]$  with $f(x)=-\frac{1}{4}\log\left[\sin^2\left(\frac{\pi}{2}x \right)\right]$.
The function $f(x)$ admits the following representation in the Fourier space (summation over integer values of the momentum $k$)
\be
f(x)=\sum_{k> 0} \frac{\cos(\pi k x)}{\pi k}\, .
\ee
Out of this repesentation, we construct a discrete approximation truncating the series
\be
f_j^d=\sum_{k> 0}^{2N_d-1} \cos\left(\frac{\pi k j}{N_d}\right)\frac{1}{\pi k}\, .
\ee
Where the index $j$ runs from $0$ to $2N_d-1$ and are a discretization of the real space. Then, we construct the discrete correlation function $[\Phi^d]_{jj'}$ as per
\be
[\Phi^d]_{jj'}=K\times [f^d_{j-j'}-f^d_{j+j'}]\, .
\ee

We now want to consider an interval $[x_1,x_2]=[i/N_d,i'/N_d]$, we then construct the discrete version of the $\Phi_\Omega$ operator as per
\be\label{eq_dis_twist}
[\Phi_\Omega^d]_{jj'}=\begin{cases} \Phi_{jj'}^d\hspace{1pc} &j\in[i,i'] \hspace{1pc} j'\in[i,i']  \\ \Phi_{jj'}^d\hspace{1pc} &j\notin[i,i'] \hspace{1pc} j'\notin[i,i'] \\
e^{-i2\pi \Omega}\Phi_{jj'}^d\hspace{1pc} &j\notin[i,i'] \hspace{1pc}j'\in[i,i']\\
e^{i2\pi \Omega}\Phi_{jj'}^d\hspace{1pc} &j\in[i,i'] \hspace{1pc} j'\notin[i,i']\end{cases}
\ee

The ratio of Fredholm determinants is discretized as per
\be\label{eq_app_discrdet}
\det\left(\frac{1+\Phi^{-1}\Phi_\Omega}{2}\right)\to \det\left(\frac{ 1+[\Phi^d]^{-1}\Phi_\Omega^d}{2}\right)\, .
\ee

The matrix $[\Phi^d]^{-1}$ has the following explicit expression
\be
[\Phi^d]^{-1}_{jj'}=\frac{1}{K N_d^2}(g^d_{j-j'}-g^d_{j+j'})\,, \hspace{3pc} g_j^d=\sum_{k> 0}^{2N_d-1} \cos\left(\frac{\pi k j}{N_d}\right)\pi k\, .
\ee

\begin{figure}[t!]
\includegraphics[height=0.21\textheight]{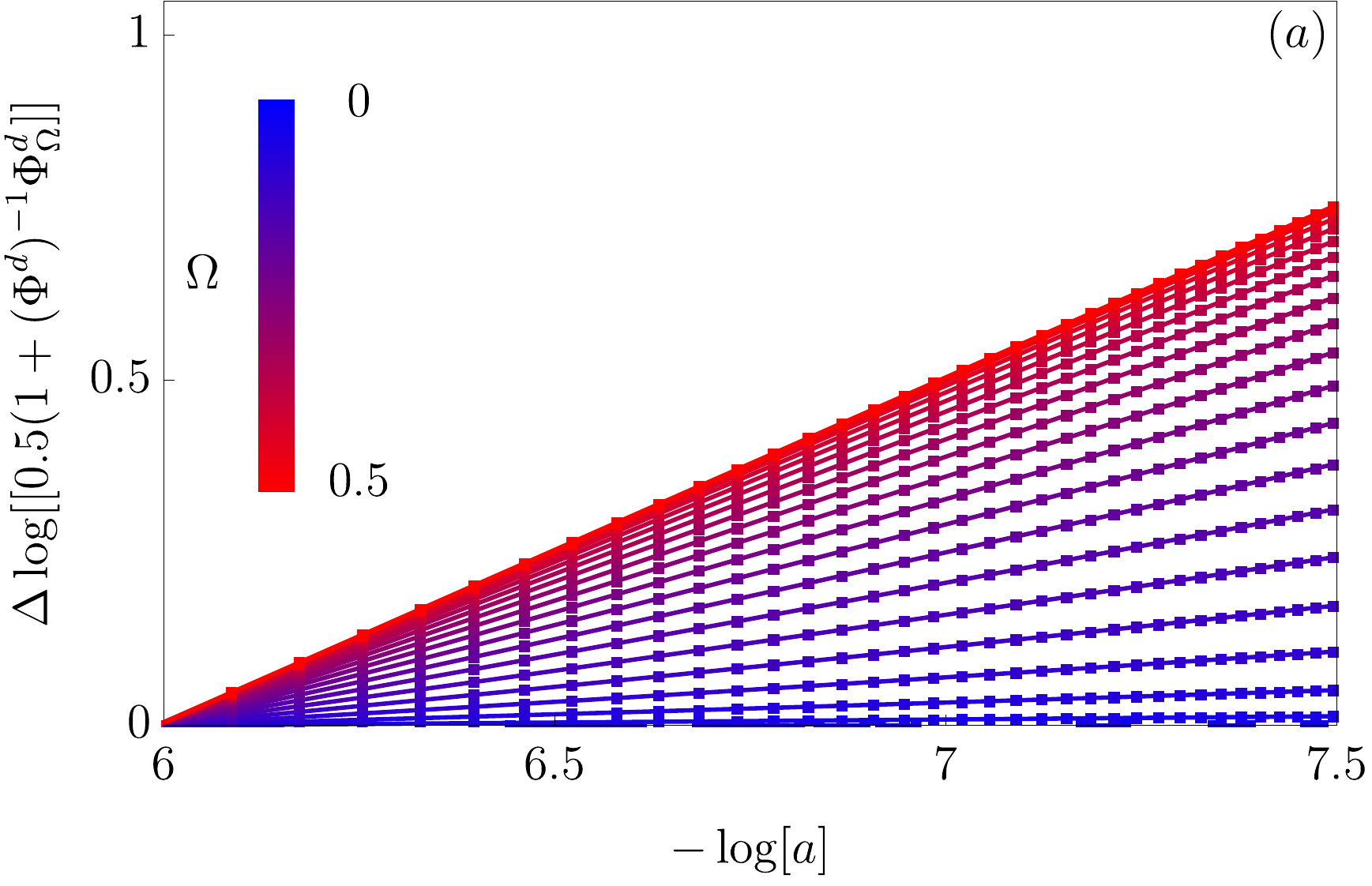}\hspace{3pc}\includegraphics[height=0.21\textheight]{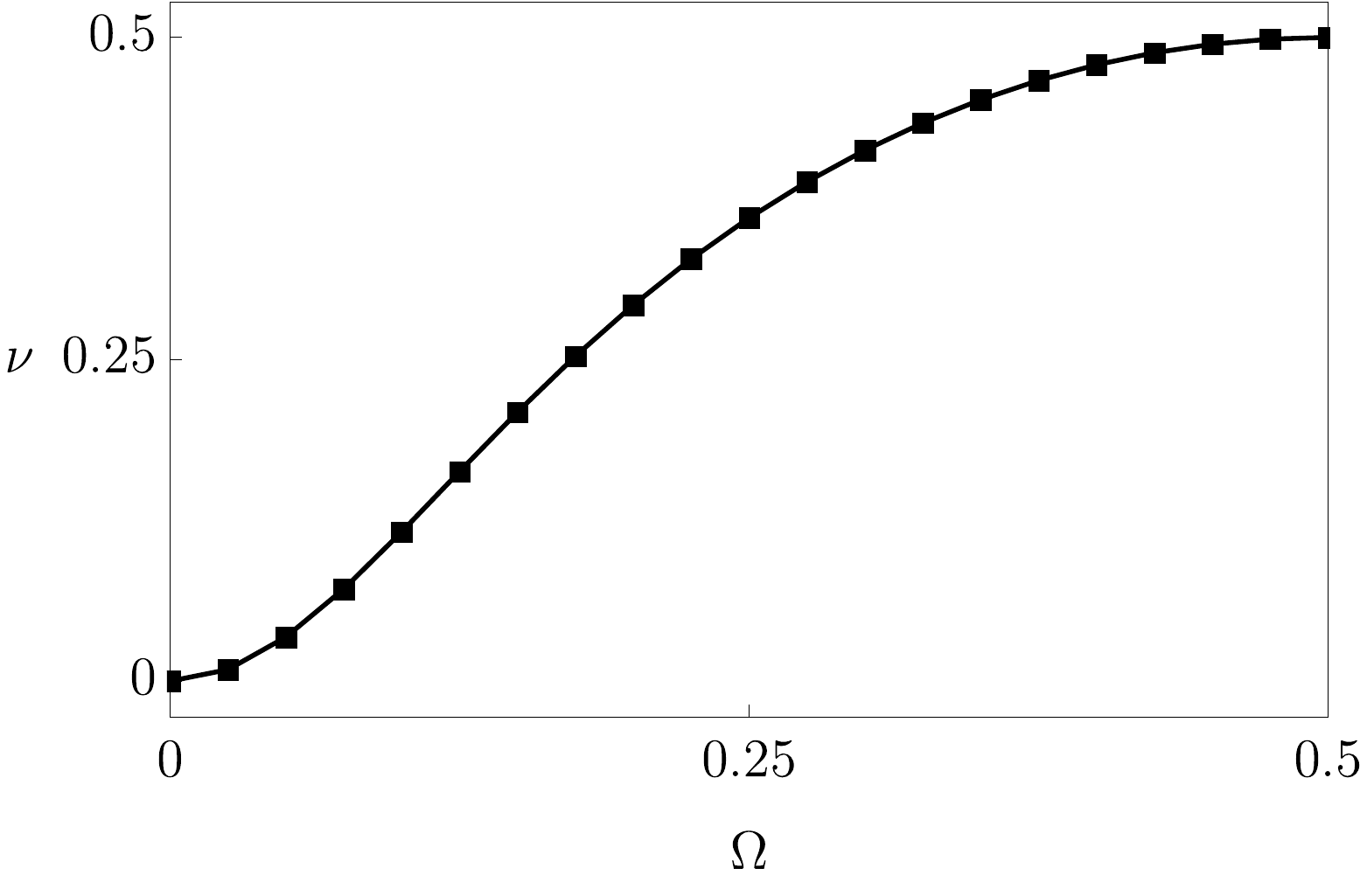}
\caption{\label{fig_sing_det}\emph{We numerically study the singularity of the operatorial determinant discretized as per Eq. \eqref{eq_app_discrdet} as a function of the lattice spacing $a$. We choose $L=1$, $K=1$ and a bipartition $A\cup B$ with $B=[0.25,0.75]$ and check the equivalence of other choices.
We progressively reduced the lattice spacing $a=1/N_d$ used in the discretization, observing a power law divergence of the determinant $\simeq a^{-\nu}$, the exponent being a non trivial function of the phase $\Omega$. 
In order to emphasize the power law behavior, in Fig. $(a)$ we consider the logarithm of the determinant as a function of $-\log(a)$ for various choices of $\Omega$. Each curve is shifted with a global offset in such a way they have a common origin: the perfect linear growth guarantees the power law behavior.
In Fig. $(b)$ we numerically extract the exponent $\nu$ and plot it as a function of $\Omega$: we focus on the case $0\le\Omega\le 0.5$, since the problem is symmetric $\Omega\to 1-\Omega$.
}}
\end{figure}

\begin{figure}[t!]
\includegraphics[height=0.21\textheight]{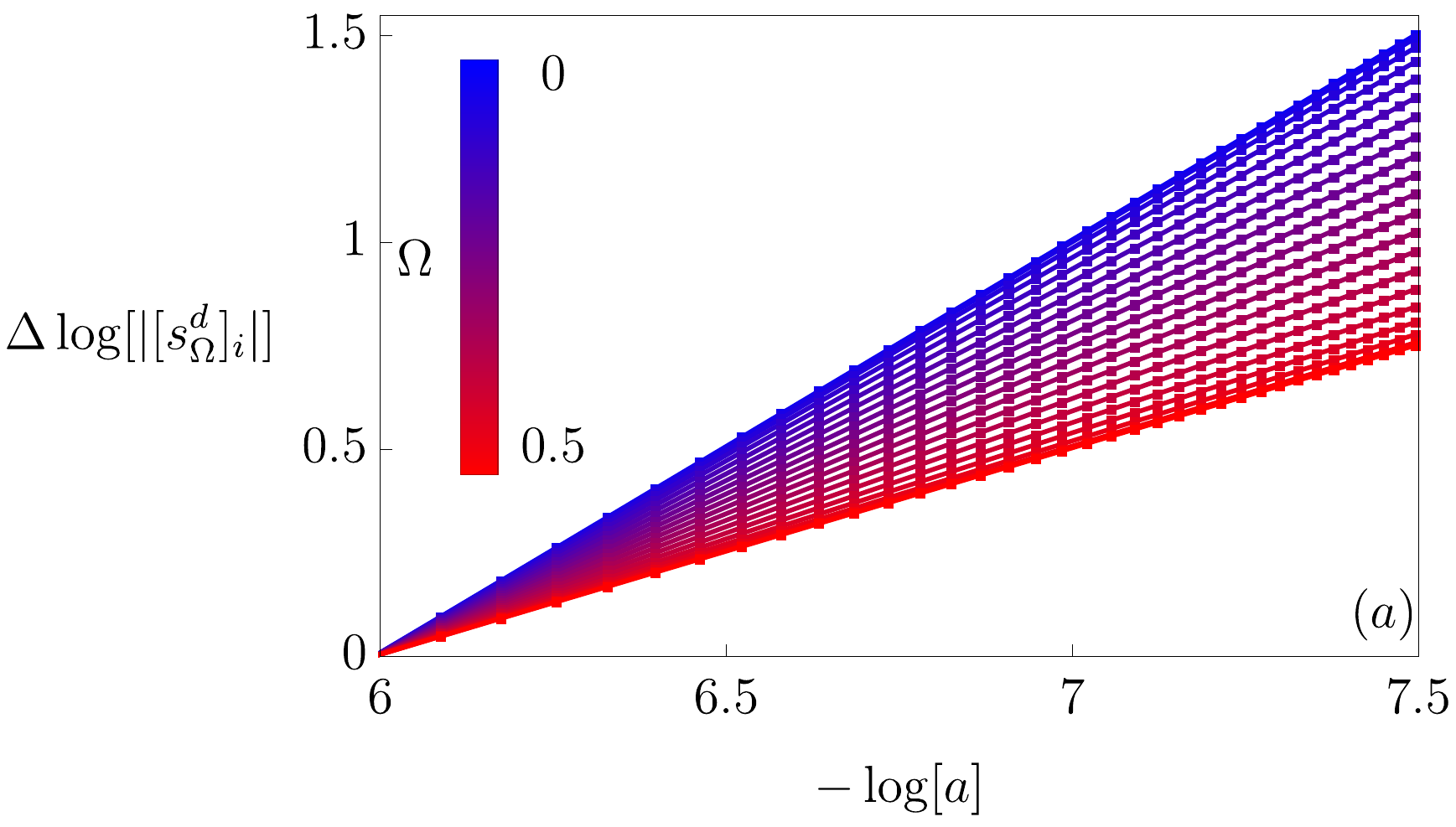}\hspace{3pc}\includegraphics[height=0.21\textheight]{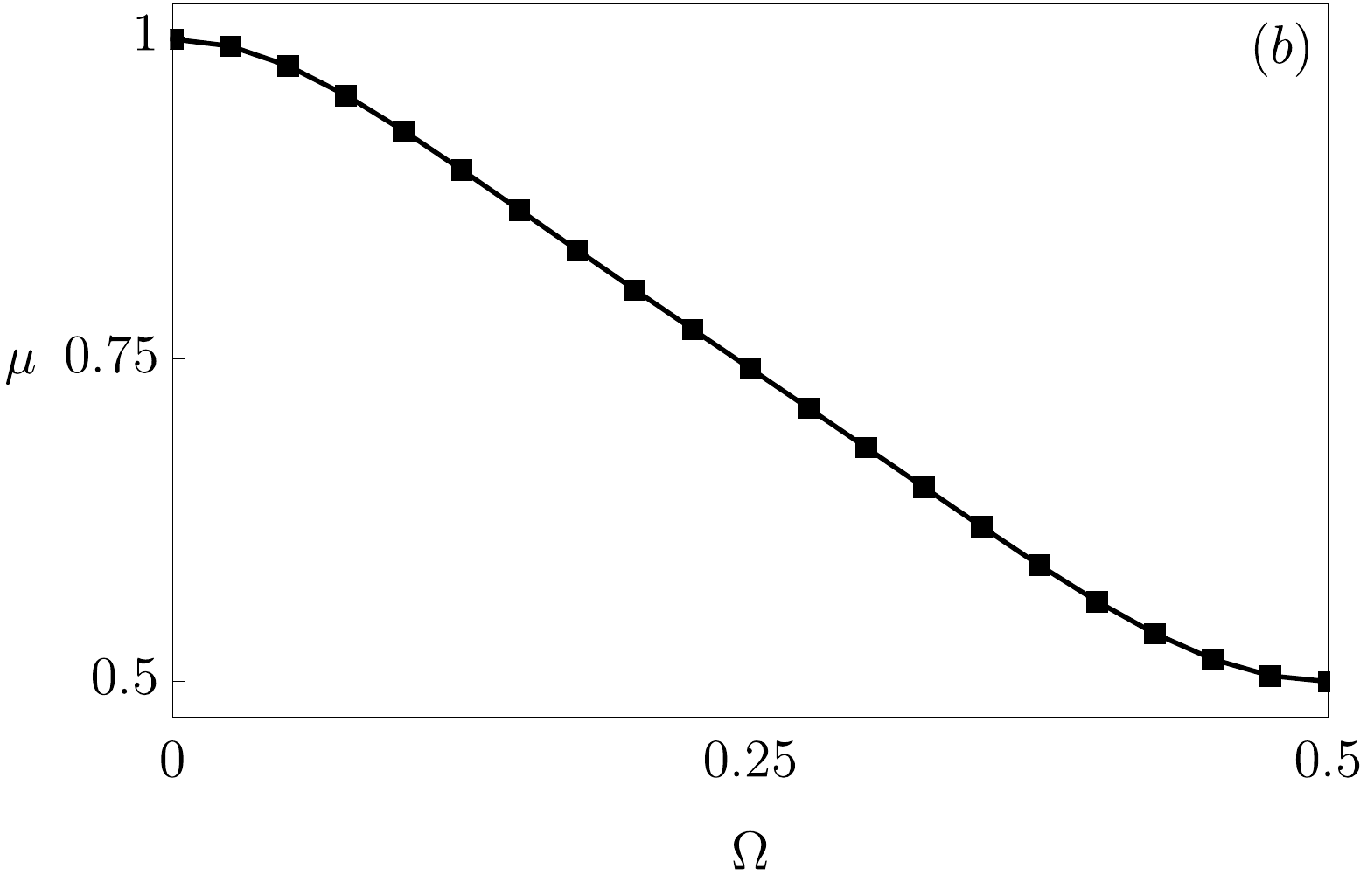}
\caption{\label{fig_sing_s}\emph{We numerically study the singularity of the solution to Eq. \eqref{eq_Il_def}, discretized according to Eq. \eqref{eq_dis_int}. As a proof of concept, we choose $L=1$, $K=1$ and a bipartition $A\cup B$ with $B=[0.25,0.75]$. Other choices of the bipartition lead to the same conlcusions. We progressively increase the number of points in the discretization $N_d$ and define $i=0.25 N_d$ as the index at which the phase twist of the left boundary is introduced, according to Eq. \eqref{eq_dis_twist}. Subfigure $(a)$: for different values of the phase twist $\Omega$, we plot the value of $\log|[s^d_\Omega]_i|$ against the lattice spacing $-\log a=\log N_d$. For the sake of clarity, rather than plotting $\log|[s_\Omega^d]_i|$, we shift each curve in such a way it starts from the origin and focus on the growth $\Delta\log|[s_\Omega^d]_i|$ with the decreasing of the lattice spacing.
The perfect linear growth testifies the power-law singularity of the solution to the continuum integral equation \eqref{eq_Il_def} $\sim |x-x_1|^{-\mu}$ (in the current example, $x_1=0.25$). The slope of the linear growth allows to numerically determine the exponent $\mu$ (Subfig. $(b)$), which we verify satisfies $0.5\le\mu\le 1$. We considered $0\le\Omega\le0.5$, since the case $0.5\le\Omega\le 1$ is trivially connected to the previous one through complex conjugation.
}}
\end{figure}

As expected, we see that the determinant has a well defined continuous limit up to a constant (i.e. it does not depend on the actual position of the interval) which diverges with the lattice spacing.
As depicted in Fig. \ref{fig_sing_det}, the determinant of Eq. \eqref{eq_app_discrdet} has a power law singularity $\propto a^{-\nu(\Omega)}$ where the exponent $\nu$ is a non trivial function of the phase twist $\Omega$. In the computation of the R\'enyi entropy one must consider the logarithm of such a determinant, resulting then in an additive constant which diverges as per $\frac{1}{2}\frac{1}{1-N}\sum_{\ell=1}^{N-1}\nu(N/\ell) \log a$.
Numerically, one can check
\be
\frac{1}{2}\frac{1}{1-N}\sum_{\ell=1}^{N-1}\nu(N/\ell)=-\frac{1+1/N}{6}\, ,
\ee
which is in agreement with the prefactor of the $\log a$ divergent term in the CFT result Eq. \eqref{eq_S_infinity}, as it should be.

We now discuss the computation of the $\mathcal{I}_{\ell/N}$ coefficients. We define the discrete characteristic function $\chi_{[i,i']}^d$ as it follows
\be
[\chi_{[i,i']}^d]_j=\begin{cases} 1\hspace{2pc}&j\in[i,i'] \\ 0\hspace{2pc}&j\notin[i,i']\end{cases}\, .
\ee
Then define a vector $[s_\Omega^d]_j$ as the solution of the linear equation below
\be\label{eq_dis_int}
a(\Phi^d+\Phi^d_\Omega)\cdot s_\Omega^d=\chi_{[i,i']}^d\, .
\ee
The presence of the lattice spacing ensures that $s_\Omega^d$ reproduces the solution of the continuum problem in the limit $a\to 0$, i.e. $[s_\Omega^d]_j\to s^\Omega(j/N_d)$. Finally, the cofficients $\mathcal{I}_\Omega$ are computed as 
\be
\mathcal{I}_\Omega=a\sum_j [s_\Omega^d]_j[\chi_{[i,i']}^d]_j\, .
\ee
The presence of the sudden phase twist in the definition of $\Phi_\Omega$ causes a power-law singularity in the solution of $s_\Omega(x)$ around $x\sim x_1$ and $x\sim x_2$. In Fig. \ref{fig_sing_s} (Subfig. $(a)$) we numerically study the singular behavior, testifying a power-law singularity $|x-x_1|^{-\mu}$. The exponent of the singularity is numerically extracted in Fig. \ref{fig_sing_s} (Subfig. $(b)$), showing the singularity is always integrable for $\Omega\ne 0\, \text{mod}\,1$ and ensuring the finitness of $\mathcal{I}_\Omega$, as wee commented in the main text.


\begin{thebibliography}{99}

\bibitem{rev-enta}
L.~Amico, R.~Fazio, A.~Osterloh, and V.~Vedral, 
Rev. Mod. Phys.  \href{https://doi.org/10.1103/RevModPhys.80.517}{\bf 80} 517 (2008).
\bibitem{rev-enta2}
P.~Calabrese, J.~Cardy, and B.~Doyon, 
J. Phys. A \href{http://dx.doi.org/10.1088/1751-8121/42/50/500301}{\bf 42} 500301 (2009).
\bibitem{rev-enta3}
N.~Laflorencie, 
Physics Report \href{http://dx.doi.org/10.1016/j.physrep.2016.06.008}{\bf 643}, 1 (2016).

\bibitem{CaCa16}P. Calabrese, J. Cardy J. Stat. Mech. (2016) \href{https://doi.org/10.1088/1742-5468/2016/06/064003}{064003}.


\bibitem{CaCa04} P. Calabrese, J. Cardy J. Stat. Mech. (2004) \href{https://doi.org/10.1088/1742-5468/2004/06/P06002}{P06002}.

\bibitem{CaCa09} P. Calabrese, J. Cardy, J. Phys. A: Math. Theor. \href{https://doi.org/10.1088/1751-8113/42/50/504005}{\bf 42} 504005 (2009).

\bibitem{HoLaWi94} C. Holzhey, F. Larsen, F. Wilczek, Nucl. Phys. B \href{https://doi.org/10.1016/0550-3213(94)90402-2}{\bf 424}, 3 443-477 (1994).

\bibitem{swvc-08}
N. Schuch, M. M. Wolf, F. Verstraete, and J. I. Cirac,
Phys. Rev. Lett. \href{http://dx.doi.org/10.1103/PhysRevLett.100.030504}{\bf 100}, 030504 (2008).
\bibitem{swvc-082}
N. Schuch, M. M. Wolf, K. G. H. Vollbrecht, and J. I. Cirac,
New J. Phys. \href{http://dx.doi.org/10.1088/1367-2630/10/3/033032}{\bf 10}, 033032 (2008).

\bibitem{pv-08}
A. Perales and G. Vidal,
Phys. Rev. A \href{http://dx.doi.org/10.1103/PhysRevA.78.042337}{\bf 78}, 042337 (2008).

\bibitem{rev0}
P. Hauke, F. M. Cucchietti, L. Tagliacozzo, I. Deutsch, and M. Lewenstein,
Rep. Prog. Phys. \href{http://dx.doi.org/10.1088/0034-4885/75/8/082401}{\bf 75} 082401 (2012).

\bibitem{d-17}
J. Dubail, 
J. Phys. A \href{http://dx.doi.org/10.1088/1751-8121/aa6f38}{\bf 50}, 234001 (2017).

\bibitem{lpb-18}
E. Leviatan, F. Pollmann, J. H. Bardarson, D. A. Huse, and E. Altman,
\href{https://arxiv.org/abs/1702.08894}{arXiv:1702.08894} (2017).

\bibitem{Ra15} R. Islam, R. Ma, P. Preiss, M. Tai, A. Lukin, M. Rispoli, M. Greiner, Nature, \href{https://doi.org/10.1038/nature15750}{\bf 528}, 77-83 (2015).

\bibitem{Kauf16} A. M. Kaufman, M. Tai, A. Lukin, A. Rispoli, R. Schittko, P. Preiss, M. Greiner, Science \href{\doi10.1126/science.aaf6725}{\bf 353}, 794–800 (2016).

%%
\bibitem{DaPi12} A. J. Daley, H. Pichler, J. Schachenmayer, P. Zoller, Phys. Rev. Let.. \href{https://doi.org/10.1103/PhysRevLett.109.020505}{\bf 109}, 020505 (2012).

\bibitem{UnZh17} J. Unmuth-Yockey, J. Zhang, P. Preiss, L.-P. Yang, S.-W. Tsai, Y. Meurice, Phys. Rev. A \href{https://doi.org/10.1103/PhysRevA.96.023603}{\bf 96}, 023603 (2017).

\bibitem{ElVe18} A. Elben, B. Vermersch, M. Dalmonte, J. I. Cirac, P. Zoller, Phys. Rev. Lett. \href{https://doi.org/10.1103/PhysRevLett.120.050406}{\bf 120}, 050406 (2018).

\bibitem{LuRi19} A. Lukin, M. Rispoli, R. Schittko, M. E. Tai, A. M. Kaufman, S. choi, V. Khemani, J. Leonard, M. Greiner, Science \href{\doi10.1126/science.aau0818}{\bf 364}, 256-260 (2019).

\bibitem{BrEl19} T. Brydges, A. Elben, P. Jurcevic, B. Vermersch, C. Maier, B. P. Lanyon, P. Zoller, R. Blatt, C. F. Roos, Science \href{\doi10.1126/science.aau4963}{\bf 364}, 260-263 (2019).


\bibitem{CaLe08} P. Calabrese, A. Lefevre, Phys. Rev. A \href{https://doi.org/10.1103/PhysRevA.78.032329}{\bf 78}, 032329 (2008).

\bibitem{AlCaTo18} V. Alba, P. Calabrese, E. Tonni, J. Phys. A \href{https://doi.org/10.1088/1751-8121/aa9365}{\bf 51}, 024001 (2018).

\bibitem{LiHa08} H. Li, F. D. M. Haldane, Phys. Rev. Lett. \href{https://doi.org/10.1103/PhysRevLett.101.010504}{\bf 101}, 010504 (2008).

\bibitem{BPZ} A. A. Belavin, A. M. Polyakov, A. B. Zamolodchikov, Nucl. Phys. B \href{https://doi.org/10.1016/0550-3213(84)90052-X}{\bf 241}, 333 (1984).

\bibitem{difrancesco} P. Di Francesco, P.  Mathieu, D. S\'en\'echal \emph{Conformal field theory} (Springer Science \& Business Media, 2012).


\bibitem{Ha81}F. Haldane, Physics Letters A \href{https://doi.org/10.1016/0375-9601(81)90049-9}{\bf 81}, 153 (1981).
\bibitem{Ha81bis} F. Haldane, Journal of Physics C: Solid State Physics \href{https://doi.org/10.1088/0022-3719/14/19/010}{\bf 14}, 2585 (1981)
\bibitem{Ha81tris} F. Haldane, Physical Review Letters \href{https://doi.org/10.1103/PhysRevLett.47.1840}{\bf 47},1840 (1981).
\bibitem{Ca04} M. Cazalilla, Journal of Physics B: Atomic, Molecular and Optical Physics \href{https://doi.org/10.1088/0953-4075/37/7/051}{\bf 37},S1 (2004).
\bibitem{Gi04} T. Giamarchi, \emph{Quantum physics in one dimension}, vol. 121 (Oxford university press, 2004).
\bibitem{Ts07} A. M. Tsvelik, \emph{Quantum field theory in condensed matter physics} (Cambridge university press, 2007).
\bibitem{CaCiGiOrRi11} M. Cazalilla, R. Citro, T. Giamarchi, E. Orignac, and M. Rigol, Reviews of Modern Physics \href{https://doi.org/10.1103/RevModPhys.83.1405}{\bf 83} 1405 (2011).


\bibitem{RuToCa18}P. Ruggiero, E. Tonni, P. Calabrese, J. Stat. Mech. (2018) \href{https://doi.org/10.1088/1742-5468/aae5a8}{113101}.

\bibitem{RaGl12} M. A. Rajabpour, F. Gliozzi, J. Stat. Mech. \href{https://doi.org/10.1088/1742-5468/2012/02/P02016}{P02016} (2012).
\bibitem{CaGl08} M. Caraglio, F. Gliozzi, JHEP 11 (2008) \href{https://doi.org/10.1088/1126-6708/2008/11/076}{076}.
\bibitem{FuPaSh09} S. Furukawa, V. Pasquier, J. Shiraishi, Phys. Rev. Lett. \href{https://doi.org/10.1103/PhysRevLett.102.170602}{\bf 102}, 170602 (2009).
\bibitem{He10}M. Headrick, Phys. Rev. D \href{https://doi.org/10.1103/PhysRevD.82.126010}{\bf 82}, 126010 (2010).
\bibitem{HeLaRo13} M. Headrick, A. Lawrence, M. Roberts, J. Stat. Mech. \href{https://doi.org/10.1088/1742-5468/2013/02/P02022}{P02022} (2013).
\bibitem{CaHu09bis} H. Casini, M. Huerta, JHEP \href{https://doi.org/10.1088/1126-6708/2009/03/048}{03} (2009) 048.
\bibitem{FaFlInPa08} P. Facchi, G. Florio, G. Parisi, S. Pascazio, Phys. Rev. A \href{https://doi.org/10.1103/PhysRevA.77.060304}{\bf 77} 060304 (2008).
\bibitem{AlTaCa10} V. Alba, L. Tagliacozzo, P. Calabrese, Phys. Rev. B \href{https://doi.org/10.1103/PhysRevB.81.060411}{\bf 81} 060411 (2010).
\bibitem{AlTaCa11} V. Alba, L. Tagliacozzo, P. Calabrese, J. Stat. Mech. (2011) \href{https://doi.org/10.1088/1742-5468/2011/06/P06012}{P06012}.
\bibitem{IgPe10} F. Igloi, I. Peschel, EPL \href{https://doi.org/10.1209/0295-5075/89/40001}{\bf 89} (2010) 40001.
\bibitem{FaCa10} M. Fagotti, P.Calabrese, J. Stat. Mech. (2010) \href{https://doi.org/10.1088/1742-5468/2010/04/P04016}{P04016}.
\bibitem{Ca10} P. Calabrese, J. Stat. Mech. (2010) \href{https://doi.org/10.1088/1742-5468/2010/09/P09013}{P09013}.
\bibitem{Fa12} M. Fagotti, EPL \href{https://doi.org/10.1209/0295-5075/97/17007}{\bf 97} (2012) 17007.
\bibitem{ChLoZh13} B. Chen, J. Zhang, JHEP \href{https://doi.org/10.1007/JHEP11(2013)164}{1311} (2013) 164.
\bibitem{ChLoZh14} B. Chen, J. Long, J. Zhang, JHEP \href{https://doi.org/10.1007/JHEP04(2014)041}{1404} (2014) 041.
\bibitem{ArEsFa14} F. Ares, J. G. Esteve, F. Falceto, Phys. Rev. A \href{https://doi.org/10.1103/PhysRevA.90.062321}{\bf 90}, 062321 (2014).
\bibitem{CoToCa16} A. Coser,E. Tonni, P. Calabrese, J. Stat. Mech. (2016) \href{https://doi.org/10.1088/1742-5468/2016/05/053109}{053109}.
\bibitem{LiZh16}Z. Li, J. Zhang, JHEP \href{https://doi.org/10.1007/JHEP05(2016)130}{1605} (2016) 130.
\bibitem{LiuLiu16} F. Liu, X. Liu, JHEP \href{https://doi.org/10.1007/JHEP01(2016)058}{01} (2016) 058.
\bibitem{BeKeZa17} A. Belin, C. A. Keller, I. G. Zadeh, J. Phys. A \href{https://doi.org/10.1088/1751-8121/aa8a11}{\bf 50}, 435401 (2017).
\bibitem{MuMu18}S. Mukhi, S. Murthy, J.-Q. Wu, JHEP \href{https://doi.org/10.1007/JHEP01(2018)005}{01} (2018) 005.

\bibitem{DuEsIk18}T. Dupic, B. Estienne, Y. Ikhlef, SciPost Phys. \href{\doi 10.21468/SciPostPhys.4.6.031}{\bf 4}, 031 (2018).
\bibitem{XaAlSi18} J. C. Xavier, F. C. Alcaraz, G. Sierra, Phys. Rev. B \href{https://doi.org/10.1103/PhysRevB.98.041106}{\bf 98}, 041106(R) (2018).




\bibitem{CaCaTo09} P. Calabrese, J. Cardy, E. Tonni,  J. Stat. Mech. (2009) \href{https://doi.org/10.1088/1742-5468/2009/11/P11001}{P11001}.

\bibitem{CaCaTo11} P. Calabrese, J. Cardy, E. Tonni, J. Stat. Mech. (2011) \href{https://doi.org/10.1088/1742-5468/2011/01/P01021}{P01021}.

\bibitem{CoTaTo14} A. Coser, L. Tagliacozzo, E. Tonni, J. Stat. Mech. (2014) \href{https://doi.org/10.1088/1742-5468/2014/01/P01008}{P01008}.


\bibitem{Ca08} J. Cardy, \href{arXiv:hep-th/0411189}{arXiv: hep-th/0411189}, (2008).




%%kondo

\bibitem{Ko64} J. Kondo,  Progr. Theoret. Phys., \href{https://doi.org/10.1143/PTP.32.37}{\bf 32} (1964), 37-49.

\bibitem{He97} A. C. Hewson, (1997). \emph{The Kondo problem to heavy fermions (Vol. 2).} Cambridge university press.

\bibitem{AfLaSo09}I. Affleck, N. Laflorencie, E. S S{\o}rensen,  J. Phys. A: Math. Theor. \href{https://doi.org/10.1088/1751-8113/42/50/504009}{\bf 42} 504009 (2009).


\bibitem{ZhBaFj06}H.-Q. Zhou, T. Barthel, J. O. Fj{\ae}restad, U. Schollw\"ock,
Phys. Rev. A \href{https://doi.org/10.1103/PhysRevA.74.050305}{\bf 74}, 050305(R) (2006).

\bibitem{AfLu91} I. Affleck, A. W. W. Ludwig, Phys. Rev. Lett. \href{https://doi.org/10.1103/PhysRevLett.67.161}{\bf 67}, 161 (1991).


\bibitem{CaFoHu05} H. Casini, C. D. Fosco, M. Huerta, J. Stat. Mech. (2005) \href{https://doi.org/10.1088/1742-5468/2005/07/P07007}{P07007}.


\bibitem{FaCa11}M. Fagotti, P. Calabrese J. Stat. Mech. (2011) \href{https://doi.org/10.1088/1742-5468/2011/01/P01017}{P01017}.

\bibitem{CaMiVi11}
P. Calabrese, M. Mintchev, E. Vicari,
Phys. Rev. Lett. \href{https://doi.org/10.1103/PhysRevLett.107.020601}{\bf 107}, 020601 (2011).
\bibitem{CaMiVi11JS}P. Calabrese, M. Mintchev, E. Vicari, J. Stat. Mech. \href{https://doi.org/10.1088/1742-5468/2011/09/P09028}{P09028} (2011).







\bibitem{ScDMRGrev}U. Schollwöck, \emph{Matrix product state algorithms: DMRG, TEBD and relatives.} In Strongly Correlated Systems (pp. 67-98). (Springer, Berlin, Heidelberg 2013).




\bibitem{CaHu09}H. Casini, M. Huerta 2009 J. Phys. A: Math. Theor. \href{https://doi.org/10.1088/1751-8113/42/50/504007}{\bf 42} 504007.


\bibitem{DuStViCa17}J. Dubail, J.-M. St\'ephan, J. Viti, P. Calabrese, SciPost Physics \href{\doi10.21468/SciPostPhys.2.1.002}{\bf 2}, 002 (2017).

\bibitem{DuStCa17} J. Dubail, J.-M. St\'ephan,  P. Calabrese, SciPost Physics \href{\doi10.21468/SciPostPhys.3.3.019}{\bf 3}, 019 (2017).

\bibitem{BrDu18}Y. Brun, J. Dubail, SciPost Phys. \href{\doi10.21468/SciPostPhys.4.6.037}{\bf 4}, 037 (2018).

\bibitem{BrDu17}Y. Brun, J. Dubail, SciPost Phys. \href{10.21468/SciPostPhys.2.2.012}{\bf 2}, 012 (2017).

\bibitem{EiBa17}V. Eisler, D. Bauernfeind,
Phys. Rev. B \href{https://doi.org/10.1103/PhysRevB.96.174301}{\bf 96}, 174301 (2017).


\bibitem{RuBrDu19} P. Ruggiero, Y. Brun, J. Dubail,  SciPost Phys. \href{\doi10.21468/SciPostPhys.6.4.051}{\bf 6}, 051 (2019)




\bibitem{MuRuCa19}S. Murciano, P.Ruggiero, P. Calabrese, J. Stat. Mech. (2019) \href{https://doi.org/10.1088/1742-5468/ab00ec}{034001}.




\bibitem{OshAf97} M. Oshikawa, I. Affleck, Nucl. Phys. B \href{https://doi.org/10.1016/S0550-3213(97)00219-8}{\bf 495} 533-582 (1997).

\bibitem{Af98}I. Affleck, J. Phys. A: Math. Gen. \href{https://doi.org/10.1088/0305-4470/31/12/003}{\bf 31} 2761 (1998).


%%numerical analytical continuation
\bibitem{AgHeJaKa14} 
C. M. Agon, M. Headrick, D. L. Jafferis, S. Kasko,
Phys. Rev. D \href{https://doi.org/10.1103/PhysRevD.89.025018}{\bf 89}, 025018 (2014).



\bibitem{DeCoTo15}
C. De Nobili, A. Coser, and E. Tonni, 
J. Stat. Mech. (2015) \href{http://dx.doi.org/10.1088/1742-5468/2015/06/P06021}{P06021}.

%entanglement negativity
\bibitem{Pe96} A. Peres, Phys. Rev. Lett. \href{https://doi.org/10.1103/PhysRevLett.77.1413}{\bf 77}, 1413 (1996).
\bibitem{ZyHo98} K. Zyczkowski, P. Horodecki, A. Sanpera, M. Lewenstein, Phys. Rev. A \href{https://doi.org/10.1103/PhysRevA.58.883}{\bf 58}, 883 (1998).
\bibitem{Zy99} K. Zyczkowski, Phys. Rev. A \href{https://doi.org/10.1103/PhysRevA.60.3496}{\bf 60}, 3496 (1999).
\bibitem{LeeKim00} J. Lee, M. S. Kim, Y. J. Park, S. Lee, J. Mod. Opt. \href{https://doi.org/10.1080/09500340008235138}{\bf 47}, 2151 (2000).
\bibitem{EiPl99} J. Eisert, M. B. Plenio, J. Mod. Opt. \href{ https://www.tandfonline.com/doi/abs/10.1080/09500349908231260}{\bf 46}, 145 (1999).
\bibitem{ViWe02} G. Vidal, R. F. Werner, Phys. Rev. A \href{https://doi.org/10.1103/PhysRevA.65.032314}{\bf 65}, 032314 (2002).
\bibitem{Pl05} M. B. Plenio, Phys. Rev. Lett. \href{https://doi.org/10.1103/PhysRevLett.95.090503}{\bf 95}, 090503 (2005).


\bibitem{CaCaTo12}P. Calabrese, J. Cardy,  E. Tonni
Phys. Rev. Lett. \href{https://doi.org/10.1103/PhysRevLett.109.130502}{\bf 109}, 130502 (2012).
\bibitem{CaCaTo13} P. Calabrese, J.Cardy, E. Tonni, J. Stat. Mech. (2013) \href{https://doi.org/10.1088/1742-5468/2013/02/P02008}{P02008}.



\end{thebibliography}
\end{document}